\documentclass{aa}
\usepackage{graphicx}
\usepackage{psfig}
\def\FeII{\ion{Fe}{ii}} 
\def\HI{\ion{H}{i}}
\def\NI{\ion{N}{i}}
\def\OI{\ion{O}{i}}
\def\SII{\ion{S}{ii}}
\def\PII{\ion{P}{ii}}
\def\SiII{\ion{Si}{ii}}
\def\CrII{\ion{Cr}{ii}}
\def\ZnII{\ion{Zn}{ii}}
\def\ArI{\ion{Ar}{i}}
\def\AlII{\ion{Al}{ii}}
\def\HII{\ion{H}{ii}}
\def\AlIII{\ion{Al}{iii}}
\def\CIV{\ion{C}{iv}}
\def\SiIV{\ion{Si}{iv}}
\def\gsim{\raisebox{-5pt}{$\;\stackrel{\textstyle >}{\sim}\;$}}
\def\lsim{\raisebox{-5pt}{$\;\stackrel{\textstyle <}{\sim}\;$}} 
\def\kms{{km s$^{-1}$}} 
\def\zabs{{$z_{\rm abs}$}}
\def\zem{{$z_{\rm em}$}} 

\def\lya{Ly\,$\alpha$}
\def\lyb{Ly\,$\beta$}
\begin{document}
\title{Early stages of Nitrogen enrichment in galaxies :\\ Clues from measurements in damped Lyman $\alpha$ 
systems\thanks{Based on observations made with the ESO 8.2 m Kueyen 
telescope operated on Paranal Observatory, (Chile). ESO programmes 65.O-0474(A) and 66.A-0594(A).}
}
%\subtitle{}
%
\author{ M. Centuri\'on\inst{1}
\and P. Molaro\inst{1}
\and G. Vladilo\inst{1}
\and C. P\'eroux\inst{1}         
\and S. A. Levshakov\inst{2} 
\and V. D'Odorico\inst{1} }
\institute{Osservatorio Astronomico di Trieste, Via G.B.Tiepolo 11, 
I-34131 Trieste, Italy \\
%\email{}
\and       
Department of Theoretical Astrophysics, Ioffe Physico-Technical Institute, 
Politechnicheskaya Str. 26, 194021 St. Petersburg, Russia     
}
\offprints{M. Centuri\'on}
\mail{centurio@ts.astro.it}
\date{}
\titlerunning{Nitrogen abundances in DLAs}
\authorrunning{Centuri\'on et al.}
\abstract{ We present   4 
new measurements of nitrogen abundances and one upper limit in damped  
\lya\   absorbers  (DLAs) obtained by means of high 
resolution (FWHM $\simeq$ 7 \kms) UVES/VLT spectra. 
In addition to these measurements we have compiled data from all DLAs 
with {measurements of} nitrogen and 
$\alpha$-capture elements (O, S or Si) available 
in the literature, including 
all   HIRES/Keck and UVES/VLT 
data for a total of 33 systems, {i.e.}
%increasing in fourteen DLAs 
the largest sample investigated so far.
We find  that [N/$\alpha$] ratios 
are distributed in two groups: 75\% of the DLAs show {a mean value of} [N/$\alpha$] {= --0.87} 
with a scatter
of {0.16} dex, while the remaining 25\%  shows ratios clustered at 
[N/$\alpha$] {= --1.45} with an even    
lower dispersion of 0.05 dex. 
The high [N/$\alpha$] 
% $\simeq$ {--0.9}  
plateau {is consistent with} the one observed 
in metal-poor \HII\ regions of blue compact   dwarf (BCD)  galaxies 
{([N/$\alpha$] = --0.73 $\pm$ 0.13)}, 
while the [N/$\alpha$] $\simeq$ {--1.5} values are  
the lowest ever observed in any astrophysical site.
{These} low [N/$\alpha$] ratios are real 
and  not due to ionization effects.   They  provide a crucial evidence 
against the primary production of N by massive stars 
as responsible for the plateau
at {--0.9/--0.7} dex observed in  DLAs and BCD galaxies.  
The transition between 
the   low-N ([N/$\alpha$] $\simeq$ --1.5)  and  
high-N ([N/$\alpha$] $\simeq$ {--0.9}) DLAs 
occurs  at a nitrogen abundance of [N/H] $\simeq -2.8$, 
suggesting that the separation may result from some peculiarity of the 
nitrogen enrichment history.
The  [N/$\alpha$] $\simeq$  --1.5  
values and {their} low dispersion are consistent with a 
modest production of primary N in massive stars; 
however, {due to the} limited sample, specially for the low-N DLAs, we cannot exclude 
a primary origin  
in intermediate mass stars as responsible for the low N abundances observed.  
}    
\maketitle           
\section{Introduction}

Nitrogen  is an important element when attempting to 
understand the chemical evolution of galaxies.
The synthesis of N in the CNO cycle during hydrogen burning is reasonably well
understood, but the characteristics of the stars which produce this element 
(range of masses, stage of evolution, etc.) are not completely clear.  
N is mostly a secondary element, produced in the CNO cycle from seed C and O 
nuclei created in
earlier generation of stars. Primary N is produced  
when freshly synthesized C in the helium-burning shell penetrates
into the hydrogen-burning shell  where it 
will be converted
in primary N by means of the CNO cycle.
In  evolution models of intermediate   mass stars, thermal 
pulses occurring during the
asymptotic giant branch AGB phase are responsible 
for the transport of the He-burning products to the H-burning shell producing
primary N (Marigo 2001).
 
For massive stars a general consensus on the transport mechanism 
is lacking, and the primary N production is strongly 
dependent on the assumed treatment
of convection in stellar interiors.      
Primary N might be
produced  in massive stars  of 
low metallicities by adjusting convective overshooting 
(Woosley \& Weaver 1995). In addition, 
stellar models which include rotation and its effects on 
the transport of elements  
show that massive stars with high rotational velocity 
produce primary N at low metallicities
(Meynet \& Maeder 2002).  
  
In  simple models of galactic evolution, 
if nitrogen has a  secondary  production, the increase of its abundance 
is proportional to the metal content of the galaxy. 
On the other hand, if nitrogen has a  primary origin,   
then its abundance is expected to increase in lockstep with that
of other primary elements.
In the classical N/O versus O/H
diagram, the secondary nitrogen production  
would be represented by a straight  line at 45$^o$ slope, while in the case of 
a primary origin of N a horizontal line is expected in this diagram 
(Talbot \& Arnett 1974).

In order to obtain clues on the 
nucleosynthetic origin of nitrogen, a considerable number of works 
has been devoted to the  
comparison of the chemical evolution model predictions with the behavior of 
the  observed N/O 
ratios as a function of metallicity  measured by oxygen abundance   
(Edmunds \& Pagel 1978; Pilyugin 1992;
Marconi et al. 1994; Izotov \& Thuan 1999; 
Henry et al. 2000; Chiappini et al. 2002; 
Calura et al. 2002; Pilyugin et al. 2002; among others).

Since determination of nitrogen abundance in stars   is  relatively 
difficult to obtain, 
the bulk of the N/O data used for this purpose are 
measurements in \HII\ emitting regions of dwarf irregulars
 (Thuan et al. 1995; Kobulnicky \& Skillman 1996; 
Izotov \& Thuan 1999)  
and spiral galaxies 
  (see Henry et al. 2000,  and Pilyugin et al. 2002,
for a compilation of data).   

In  metal-poor  \HII\ regions of dwarf galaxies
([O/H]\footnote{Using the customary definition 
[X/H] = log (X/H) -- log (X/H)$_{\sun}$. 
}
$\lsim$ --1),  
N/O ratios are roughly constant   and independent of metallicity --- 
a behavior which is interpreted as an evidence for  primary 
production of N.  
However, at higher metallicities, N/O ratios increase with O/H,  
suggesting the secondary origin of N.
In the low metallicity regime the observed N/O plateau in 
blue compact dwarf   (BCD) galaxies, 
has been interpreted as either due to the primary production of 
N by massive stars (Izotov \& Thuan 1999, Izotov et al. 2001) 
or due to primary production of N
by intermediate mass stars (Henry et al. 2000; Chiappini et al. 2002).

Determinations of N abundance in sites of low metallicities play 
an important role
in understanding the origin of N production. In this framework  
N measurements in high redshift   damped \lya\ absorbers (DLAs)  --- 
 the
  quasar  absorbers with the highest neutral hydrogen column 
  densities 
$N$(\HI) $> 2\times10^{20}$ cm$^{-2}$
(Wolfe et al. 1986) ---  are of great importance 
since their metallicities 
extend over a much lower range than those of  metal-poor 
\HII\ regions in dwarf galaxies,  
allowing to probe the  early
phases of galactic chemical enrichment.

After the very first studies  of nitrogen in DLAs  
(Pettini et al.1995; Green et al. 1995;
Vladilo et al. 1995), 
systematic measurements and compilations of extant data
have been presented by Lu et al. (1998), Centuri\'on et al. (1998) and, 
more recently, by Pettini et al. (2002) and Prochaska et al. (2002).  
In these two latter works a 
total of 10 and 19  nitrogen measurements have been presented, respectively.
Only in the data base of Pettini and collaborators, oxygen abundances  
have been considered, and are available for 6 DLAs.
Oxygen measurements, which are difficult to obtain, 
have been often replaced by measurements
of other $\alpha$-capture elements for the purpose of investigating
trends of the nitrogen abundances. 
N/O (or N/$\alpha$) ratios in DLAs show, in general 
(see, however, Prochaska et al. 2002), a large scatter, 
at variance with the near to constant 
value observed in  BCD galaxies  
at comparable metallicities. 

The N/O scatter 
observed in DLAs  has commonly been ascribed to the time delay  
between the release of O  
by short-lived, massive stars and the release of N by intermediate 
mass stars,  
which takes place over larger time scales 
 (Lu et al. 1998; Pettini et al. 2002).  
However,  as pointed  out by Centuri\'on et al. (1998),
it is difficult to conciliate
this interpretation with the absence of the expected 
enhancement of the $\alpha$-elements over the Fe-peak elements 
in the 
lowest N/$\alpha$ DLAs.
In order to probe the validity of the time delay interpretation, 
it is therefore important to add further
observational constraints based on the relative abundances of other elements. 

In this paper we describe new abundance determinations in  6  DLAs 
located in the direction of 4 QSOs. It  was  possible to 
 determine  the N abundance in only 4 of these systems.   
The measurements are generally based on 
the six lines of the \NI\ multiplets at rest wavelengths 
$\lambda\lambda$ 1134 and 1200 \AA\ 
redshifted to the optical spectral region (the redshifts of the absorbers 
lie in the range 1.9 to 2.8). 
Our results, together with the rest of 
 the published  data, are discussed in
the framework of the time delay model of chemical evolution. We compare the 
trends of N/$\alpha$ ratios  with  $\alpha$/Fe --- 
a well known indicator of chemical evolution in galaxies. 
The comparison between the trends observed in different abundance ratios
should yield significant constraints on the nucleosynthetic processes at
work in DLAs and, eventually, on the nature of the associated galaxies.   

The paper is organized as follows. 
In  Sects.  2 and 3 we present our observations and  column
density measurements,  respectively. 
In addition to N we present new measurements
of the $\alpha$-capture elements  S and Si 
%#
%#
%#
as well as of the iron-peak elements Fe and Zn.    
A compilation of 
abundance measurements for all  DLAs  
with nitrogen detections, based on our data and on data from the literature,
is presented in  Sect.  4. 
The behavior of the  ratios N/$\alpha$  and $\alpha$/iron-peak   
as well as the 
implications for understanding the nature of  DLAs 
are discussed in  Sect.  5.
 We draw our conclusions in Sect.~6.

\section{Observations and Data Reduction}

Information on the targets and on the observations is given in Table 1.
The quasars PKS\,0528--250 and HE\,0940--1050 were observed  in the framework
of our own programmes for the study of chemical abundances at high redshifts, 
while  the data for  QSO 0841+129 and QSO 1232+0815 were taken from the UVES 
archive.
Spectra of these QSOs were obtained with the Ultraviolet-Visual 
Echelle Spectrograph  (UVES; see Dekker et al. 2000) 
on the Nasmyth focus of the 8.2m Kueyen telescope, 
second unit of the VLT at Paranal, Chile.
For each target repeated exposures were taken in order to attain the 
spectral coverage of interest and to increase the final signal-to-noise ratio. 
A total of 93 spectra were collected and analyzed. 

All the observations were carried out using dichroic filters  
to simultaneously observe two 
spectral ranges with the two arms of the spectrograph.
Different setups allowed us to     
cover the spectral ranges listed in the $8^{th}$ column of Table 1.
The slit widths of both arms of UVES were set to 1 arcsec and the CCDs were 
read out in 2 x 2 binned pixels. The full width half maximum  of 
the instrumental profile ($\Delta \lambda_{\rm instr}$), measured
from the emission lines of the arcs,  gives a mean resolving power of 
R = $\lambda$/$\Delta \lambda_{\rm instr}$  $\simeq$ 42500 corresponding to a 
velocity resolution of 7 \kms.
  
The data reduction was performed using the ECHELLE context routines 
implemented in the ESO MIDAS package. 
The flat-fielding, cosmic-ray removal, sky 
subtraction, extraction, and wavelength calibration 
were performed separately on the
different spectra of each QSO. Internal errors in the wavelength calibrations 
lie in the range from 0.6 to 0.9 m\AA. After calibration, 
the observed wavelength scale was  
transformed into a vacuum, heliocentric wavelength scale. 
At this point the spectra of each QSO
with same exposure time and spectral coverage were  
averaged using the continuum level  as  a weighting factor. 
The spectra of PKS 0528-250,  obtained in different observing runs  
and with overlapping spectral regions issued from different spectrograph
configurations, 
were added using the signal-to-noise ratio, S/N, as a weighting factor. 
 
Finally, for each spectral range under 
study the local continuum was determined
in the average spectrum by using a spline function to smoothly connect the  
regions free from absorption features.  
The final spectrum used for the analysis
was obtained by normalizing the average spectrum to these continua. 
The signal-to-noise ratios  of the  final spectra,
estimated from the $rms$ scatter of the continuum near the absorptions 
under study are given in the last column of Table 1.

%%%%%%%%%%%%%%%%% Table 1 
\begin{table*}
\caption{Journal of observations}
%\begin{center}
\begin{tabular}{lcccccccc}
\hline
\hline
QSO  & V & \zem & \zabs & date & Exp. time & Number    & Coverage   &  S/N \\
     & Mag  &      &       & d/m/y &  (s)      & of spectra&    (\AA)   &  \\
\hline
PKS\,0528--250&17.7&2.765&2.141& 2--4/2/2001&33930 & 6 &  3000--3739.5 &  16\\
   & &    & 2.811 & 6,10,12/2/2001  & 45240   & 8 &  3739.5--3864 &  41 \\
   & &    &       & 12,14/3/2001    & 45240   & 8	&  3739.5--3864 & 31 \\
   & &    &       &                 & 39585   & 7	&  3864--4522  &  31\\
   & &    &       &                 & 11310   & 2 &  4522--4790  &  30 \\
   & &    &       &                 & 46560   & 9	&  4790--4983  &  37 \\
   & &    &       &                 & 35250   & 7 &  4983--5764  &  47 \\
   & &    &       &                 & 35250   & 7	&  5846--6715  &  77\\
   & &    &       &                 & 46560   & 9	&  6715--6813  &  100\\
   & &    &       &                 & 11310   & 2	&  6813--8530  &  36\\
   & &    &       &                 & 11310   & 2 &  8677--10400 &  36\\

0841+129 & 17.0 & 2.200& 2.375& 29,30/4/2000  & 5400  & 2  &3453--4776  & 10\\
  & & & 2.476&         & 5400  & 2  &6715--8530; 8677--10400  & 20--15 \\

HE\,0940--1050&16.6&3.054&1.917&26--29/3/2000&13500 &4 &3300--3864  & 25\\
 & & &    & 3/4/2000      &       &4   &4790--5763; 5763--5846  & 45--35 \\
 & & &      &               & 13500 &4   &3740--4983  & 19 \\
& & &      &               &       &4   &6715--8530; 8677--10400  & 40--35\\
1232+0815 &18.4&2.567&2.337&6,8/4/2000 & 10800 &3   &3260--4555  &   10--20\\
 & & & &          & 18000 &3   &4584--5644; 5644--6686  & 17--15 \\
\hline
\\
\end{tabular}
\\
\scriptsize{The slit was $1''$ and the CCD binning $2\times2$ 
for all the observations.\\
Dates given in Table 1 correspond to the complete set of data for 
a single QSO.\\
Intervals in the dates indicate that the spectra were obtained 
during all nights included in the interval.} 
\end{table*}

\section{Column densities}

Column densities were obtained by fitting theoretical Voigt profiles
to the observed absorption lines via $\chi^2$ minimization. This was done
using the routines FITLYMAN (Fontana \& Ballester 1995) included in the MIDAS
package. To reproduce the observed profiles,
the theoretical profiles are convolved with the 
instrumental point-spread function modeled using the emission lines
of the arcs. Portions of the profiles recognized as contaminated by intervening
\lya\ clouds were excluded from our analysis. 
The FITLYMAN routines determine the redshifts, column densities, and
broadening parameters ($b$ values) of the
absorption components, as well as the formal fit errors for
each   of these quantities. 
The laboratory wavelengths of the transitions investigated are listed 
in Table 2 together with the oscillator 
strengths adopted for the computation of
the theoretical profiles.

In the case where the absorber was fitted with a single component and several 
unsaturated transitions were detected 
for a given ion, the column density was
estimated by applying the fit procedure both to the individual
transitions and to the full set of available lines. 
No significant differences were found between the column densities 
of individual fits and the column density obtained from the 
simultaneous fit of all the lines. 
In these cases we adopted the dispersion of individual measurements as
the estimate of the  column density error.  
When a metal absorption was fitted with multiple components,
errors in the total column density were estimated by 
taking the difference between the value obtained from the fit and a  
"maximum" and a "minimum" total column density computed 
by adding the   $N$(X)$+ 1\sigma$ and  $N$(X)$- 1\sigma$, respectively, of each
component.

For saturated transitions, lower limits to the column density were estimated
from the equivalent width obtained from the best fit to the absorption.
For undetected transitions, upper limits  
were derived from  3$\sigma$ upper limits of the equivalent width.
In both cases the conversion from equivalent widths to column
densities was performed in the  optically thin regime (linear part of the
curve of growth). 

Nitrogen column densities were determined studying the
six transitions of the two \NI\ multiplets at
$\lambda$$\lambda$ 1134 and 1200 \AA\ (see Table 2). 
These lines are located in the Ly $\alpha$ forest, but 
the identification of the absorptions is  
reliable because it is unlikely that
each one of the six  transitions is   blended with a Ly $\alpha$
interloper.  The oscillator strengths span about one order of
magnitude offering a large dynamical range for the measurement  
of the column density.  
The detection of the faintest transitions of the $\lambda$ 1134 \AA\
multiplet is particularly important to avoid saturation effects.
 
Comparison of N abundance with the one of other elements 
is necessary in order to give an insight into the nucleosynthetic 
origin of nitrogen.
We therefore searched  for transitions of important species 
such as \OI, \SiII, \SII\ representative of $\alpha$-capture elements,
as well as \FeII, and \ZnII, representative of Fe-peak elements. 
The  comparison of Zn and Fe abundances is also important to assess
possible effects of dust depletion  (Vladilo 1998, 2002a).  

Oxygen is particularly important as a reference for nitrogen abundances.
As nitrogen,
this element is believed to be essentially undepleted onto dust grains.
Unfortunately,
accurate oxygen column density estimates are rare in DLAs since 
traditionally estimates are based on the strongly
saturated  \OI\ 1302.2 \AA\ absorption and on the extremely 
weak \OI\ 1355.6 \AA\ line
(never detected). 
These measurements yield  lower and upper limits, respectively,
usually differing by over one order of magnitude. 
Only recently, and mainly thanks to the high resolution and  the 
ultraviolet-visual coverage of 
the UVES  spectrograph, 
accurate \OI\ column densities  
have been obtained in DLAs by using \OI\ lines  which fall     
shortward of the \lya\ absorption and have a large
dynamical range in oscillator strengths 
 (Lopez 1999; Molaro et al. 2000, 2001; 
Dessauges-Zavadsky et al. 2001; Prochaska et al. 2001; Levshakov et al. 2002;
Pettini et al. 2002; Prochaska et al. 2002).  
Unfortunately, the \OI\ transitions 
at $\lambda$ $<$ 1302 \AA\
have not been detected for any of the DLAs studied here.  

In order to estimate the absolute abundances of the elements,  
we also derived the \HI\ column densities of the systems by fitting
the  \lya\ absorptions and possibly other lines of the Lyman series, as shown in Fig.~1. 
The resulting column densities  and abundances 
($\log N$(X), [X/H])  are given in 
Tables 3 to 8. 
Through this paper we use N and O phostospheric solar values given by 
Holweger (2001), and 
for the remaining elements the meteorite values  from 
Grevesse \& Sauval (1998).

More details on 
the measurements performed in individual systems are given in the 
following sections.

\subsection{DLAs  towards PKS 0528--250}
%at \zabs=2.141 and \zabs=2.811

PKS 0528-250 has an emission redshift of \zem =
2.779  determined from \CIV\ and \SiIV\ emission lines
(Foltz et al. 1988). 
Due to the presence of a damped \lya\ absorption at  
\zabs=2.811, the \lya\ emission of the quasar 
is partially absorbed. 
There is an additional  DLA at  \zabs=2.141 in the spectrum 
of this quasar. 
Metal abundances in both DLAs have
been previously studied by Lu et al. (1996) using HIRES/Keck spectra.
Our UVES/VLT spectra have higher S/N and extended wavelength coverage
(see Table~1).  
By means of these UVES data a study of the  molecular hydrogen  H$_2$   
in the \zabs=2.811 DLA  has  been presented in  
 Levshakov et al. (2003a).  
The detailed study of  metal abundances of both DLAs 
is given in P\'eroux
et al. (2003).  Here we present the total column densities of
nitrogen, $\alpha$-capture elements (S, Si) and Fe-peak element (Fe, Zn),
in order to discuss the abundance ratios involving nitrogen.

\subsubsection{The \zabs=2.1410 system}

The Ly$\alpha$ profile of the DLA is used to determine the neutral
hydrogen column density, $\log N$(\HI) = 20.95$\pm$0.05. The fit to
the observed profile is shown in Fig.~1. No other absorptions of the
Lyman series are observed since the quasar flux is
completely absorbed at $\lambda_{obs} < $ 3500 \AA.  
Our derived \HI\ column
density is 0.25 dex larger than the one estimated by Morton (1980) and
adopted by Lu et al. (1996).

The metal lines associated with  this system are
fitted with 10 components. In this case,
we use the numerous \FeII\ lines to fix the redshift $z$ and Doppler
parameter $b$ of each component. The derived $z$ and $b$ values are
further used to fit \NI, \SiII, and \SII.  The fitted profiles are
shown in Fig.~2 and the resulting total column
densities are given in Table~3.

The nitrogen abundance is derived for the first
time in this system. 
Only two lines out of the two triplets are suitable for column
density determination:  $\lambda\lambda$ 1134.4 and 1134.9 \AA.  
The \NI\ triplet around 1200 \AA\ is totally blended. 
The resulting fits are
shown in Fig.~2, corresponding to a total column
density $\log N$(\NI) = 14.58$\pm$0.08.

The \OI\ abundance cannot be derived  since the 1302
\AA\ line is heavily saturated and blended. On the contrary, we can
derive a reliable \SiII\ column density measurement from the 1808 \AA\
transition. The \SiII\ 1304 and 1526 \AA\ lines are heavily saturated. We
obtain a total column density of 
$\log N$(\SiII) = 15.22$\pm$0.05,
slightly lower than the one derived by Lu et al. (1996).  In addition,
we derive the column density of \SII\ from the 1253 \AA\ line, since 
the \SII\ 1250, and 1259 \AA\
transitions are completely blended. 
The last components fitted in \FeII\ and other elements are too
weak to be detected in \SII\ 1253 \AA. 
We derive for the first time the \SII\ column
density in this system obtaining 
$\log N$(\SII) = 14.83$\pm$0.04. 
Several \FeII\ lines are at our disposal to derive the column
density of this element. We use the $\lambda\lambda$ 1125, 1608, 1611,
2344, 2374, and 2586 \AA\ to deduce a total column density of
$\log N$(\FeII) = 14.85$\pm$0.09. Taking into account differences in
the adopted oscillator strengths, this result is in agreement with the one
of Lu et al. (1996). Finally, \ZnII\ 2026 \AA\ is heavily blended with 
\AlII\ 1670 \AA\  from the DLA  at \zabs=2.812. 
At the expected position for \ZnII\ 2062 \AA\ 
in the \zabs=2.141 system, we observe a weak absorption extended 
over the    
three strongest components detected in the other metals. However,   
extra absorptions are located  at both  sides of this feature due to the 
\CrII\ 2062 \AA\  and other unidentified lines (see Fig.~2).
In this case
we conservatively adopt the result of the fit as
an upper limit since we cannot exclude that the fitted 
absorption is also contaminated.

\subsubsection{The  \zabs=2.8120 system}

In our UVES spectra,  
several lines down the Lyman series are available for
determination of neutral hydrogen column density. 
We refer to Levshakov et al. (2003a) 
where a simultaneous analysis of the whole \lyb\ and the
red wings of the Ly\,$\gamma$, Ly\,5 -- Ly\,8 leads to the column
density determination: $\log N$(\HI) = 21.11$\pm$0.04.

The velocity profiles of the metals are extremely complex but well
fitted with 16 components for most species as it can be seen from
Fig.~3. We constrain the number of components by
fitting simultaneously the profiles of \SII\ 1250 \AA\ and 1253 \AA\
absorptions and \ArI\ 1048 and \PII\ 1152
(see P\'eroux et al. 2003 and Vladilo et al. 2003). 
As in the case of the 
\zabs=2.141 DLA, we then fix the redshift $z$ and Doppler parameter
$b$ of each component using \SII\ and \ArI\ 
to be equal for all elements. The resulting
total column densities are presented in Table~4.

We determine a conservative upper limit to the \NI\ column density by
considering the most constraining regions of the profiles of \NI\
953.4 \AA, together with the $\lambda\lambda$ 1134 and 1200 \AA\ multiplets. 
In this way we estimate the 
maximum \NI\ column density 
$\log N$(\NI) $\le$ 15.66, consistent with
the observed profiles (see upper panel of Fig.~3).
 
Unfortunately, the column densities of \OI\  can not be
determined in this absorber due to saturation and blending of the
lines. However, accurate column densities of \SII\ have been derived
thanks to the absorptions of two different transitions ($\lambda\lambda$
1250 and 1253 \AA). The resulting total column density is 
$\log N$(\SII) = 15.56$\pm$0.02. 
For \SiII\ only the $\lambda$ 1808 \AA\ transition is not 
heavily saturated, but it is
contaminated by few atmospheric absorption lines. 
We use the spectrum of a 
fast rotator standard star to correct from the telluric absorptions 
and we obtain
$\log N$(\SiII) = 16.01$\pm$0.03 
in perfect agreement with the value reported by Lu et al. (1996). 
Five \FeII\ lines are used to measure the \FeII\
abundance: $\lambda\lambda$ 1081, 1096, 1125, 1608 and 2374 \AA\ leading to
$\log N$(\FeII) = 15.47$\pm$0.02. 
Finally, the \ZnII\ column density
is estimated only from the transition $\lambda$ 2026 \AA, since
$\lambda$ 2062 \AA\ is blended, giving 
$\log N$(\ZnII) = 13.27$\pm$0.03. 
This column density is significantly larger
than the value 
$\log N$(\ZnII) = 13.09$\pm$0.07 given by Lu et al. (1996) and the  difference  is likely
due to the different tracement of the continuum.

\subsection{DLAs towards QSO 0841+129}
%at \zabs=2.3745 and \zabs=2.4762 
 
The two damped systems in the spectrum of QSO 0841+129 have been 
previously studied at high resolution by  Prochaska \& Wolfe  (1999) 
with HIRES/Keck spectra,
by Pettini et al. (1997) and Centuri\'on et al. (2000)
with ISIS/WHT spectra of much lower resolution 
(R $\simeq$ 2500 and 5000, respectively).
Here we present the first \NI\  measurements  obtained for these systems
and much more accurate column density determinations of \SII\ 
(in both systems)  
and \FeII\ (in the bluest one),  than previously
determined  from low resolution data (Centuri\'on et al. 2000).

\subsubsection{The  \zabs=2.3745 system}

Column densities and  metal  abundances are listed in Table~5.
Metal lines from which abundances are  obtained are shown in Fig.~4.

The \lya\ absorption is used to derive the \HI\ column density,
$\log N$(\HI) = 21.00$\pm$0.10, 
which is in agreement with the value
$\log N$(\HI) = 20.95$\pm$0.10 
previously published by
Pettini et al. (1997), and Centuri\'on et al. (2000).   

For \NI\ we restrict  the analysis to the $\lambda$$\lambda$ 
1134.4, 1134.9 \AA\ lines, 
the only unsaturated and uncontaminated ones, and we obtain 
$\log N$(\NI) = 14.62$\pm$0.03.
In Fig.~4 we show the synthetic spectrum built  
with the parameters obtained from the fit of these two \NI\ transitions.  
The synthetic spectrum is in
excellent agreement with the observed  profiles of all the 
six \NI\ absorptions.

The \SiII\ transitions $\lambda\,\lambda$ 1190,1193,1260,1304 \AA\ 
present in our 
spectrum are heavily saturated. The most stringent lower limit we can obtain
from the weakest $\lambda$ 1304 \AA\  transition is 
$\log N$(\SiII) $\geq$ 14.65.
This is consistent with the measurement from Prochaska et al. (1999),
$\log N$(\SiII) = 15.24$\pm$0.03, obtained  
from the unsaturated \SiII\, 1808 \AA\ transition, 
that we adopt in the rest of this paper.     

The \SII\ column density  
is obtained from the \SII\, 1259 \AA\, transition which lies 
in the red wing of the  \zabs = 2.4762  \lya\ absorption. 
The presence of
this \lya\ profile precludes the detection of the
other two bluer absorptions of the  \zabs = 2.3745 \SII\ triplet. 
The present result, 
$\log N$(\SII) = 14.77$\pm$0.03, is 
one order of magnitude more accurate than
the one previously obtained from ISIS/WHT data, 
$\log N$(\SII) = 14.92$^{+0.16}_{-0.21}$. 
  
The \FeII\ transitions used to obtain the iron column density
are shown in Fig.~4.
The redder \FeII\ transitions at $\lambda \lambda$ 2260, 2344, 2374, 2382 \AA,
not shown in the figure, are all saturated. The derived column density, 
$\log N$(\FeII) = 14.87$\pm$0.04 is in perfect agreement 
with the one obtained from ISIS/WHT spectra, 
$\log N$(\FeII) = 14.83$\pm$0.15,
but again  has an order of magnitude higher accuracy.  

The \ZnII\ column density obtained here from the $\lambda$ 2026 \AA\ 
transition ---  
$\log N$(\ZnII) = 12.20$\pm0.05$  --- is in agreement, within the errors, with 
the result 
$\log N$(\ZnII) = 12.12$\pm0.05$ obtained by Prochaska \& Wolfe (1999). 
We can not use
the \ZnII\ $\lambda$2062 \AA\ absorption because it is contaminated by 
a sky emission line. 
The [Zn/Fe] abundance ratio are about solar, 
indicating the absence of dust depletion
in this system  (see Vladilo 2002a, and references therein).

\subsubsection{The  \zabs=2.4762 system} 
Column densities and  metal 
abundances in this system are given in Table~6. 
The 
metal absorption lines from which elemental abundances 
are derived are shown in Fig.~5.

The damped \lya\ and \lyb\ absorptions are used to derive 
the \HI\ column density,
$\log N$(\HI) = 20.78$\pm$0.08, 
in perfect agreement with 
$\log N$(\HI) = 20.79$\pm$0.10
reported by Pettini et al. (1997).

The \NI\ column density is obtained  from 
the analysis of the $\lambda\lambda$ 1134.2, 1200.2 and 1200.7 \AA\ 
transitions.
The resulting parameters are used to  build   
the synthetic spectrum for the six \NI\ transitions shown in  Fig.~5.  
 
The \SiII\ transitions $\lambda$$\lambda$ 1190, 1193, 1260, 1304 \AA\ 
are saturated. The weakest 
transition 1020 \AA\ has an asymmetric  red   wing 
which is not observed in any other feature of this absorber. 
In this case we fix the central wavelength at the redshift  observed in the 
\SII\ triplet and  we obtain  
$\log N$(\SiII) = 14.95$\pm$0.10. 
Prochaska \& Wolfe (1999) conservatively adopt  the lower limit   
$\log N$(\SiII) $>$ 14.46, obtained from the saturated 1526 \AA\ line, 
arguing that   the unsaturated \SiII\,1808 line 
is blended with the \AlIII\,1862 line 
of the  \zabs = 2.3745 system. 
However, these two transitions are separated 
by about 1 \AA\ and they 
appear to be resolved in the spectrum
of Prochaska \& Wolfe (1999).
We therefore adopt our \SiII\ column density, which is in perfect agreement  
with the result obtained by Prochaska \& Wolfe (1999) from the 1808 \AA\ line, 
$\log N$(\SiII) = 14.96$\pm$0.02.
 
In the UVES spectrum, 
the lines of the sulphur triplet are clear from blending with other
absorptions present in the spectrum.
The  derived column density, 
$\log N$(\SII) = 14.59$\pm$0.03
is in agreement, within the errors, but much more accurate than the 
$\log N$(\SII) = 14.81$\pm$0.21 value
obtained from the ISIS/WHT spectrum, where the \SII\ 
triplet was not fully resolved. 
  
The \FeII\ column density, 
$\log N$(\FeII) = 14.55$\pm$0.07,  
is based on the analysis 
of the unsaturated and  uncontaminated   
$\lambda\lambda$ 1081 and 1096 \AA\ transitions. Our value is
consistent within the errors with the range
of values 14.52 $<$ $\log N$(\FeII) $<$ 14.54
obtained by Prochaska \& Wolfe (1999)
from the saturated $\lambda$ 1608 \AA\ and the undetected 
$\lambda$ 1611 \AA\ transitions.
Prochaska \& Wolfe (1999)   estimated   a  
more stringent upper limit to the \ZnII\ column density than the one 
obtained here (see Table~6). 
The abundances of \ZnII\ and \CrII\ derived  by  Prochaska \& Wolfe,
[Zn/H] $<$ --1.67 and [Cr/H] = --1.63, 
compared with that of iron obtained here, 
[Fe/H] = --1.69,  indicate the absence of depletion 
of the refractory elements onto dust grains.

\subsection{DLA at \zabs=1.9184 towards HE 0940$-$1050}

The Ly$\alpha$ profile of this absorber is located blueward of a
strong  Lyman limit system (LLS)  at \zabs = 2.917, in a part of
the spectrum which is thus strongly absorbed. A detailed analysis of
this LLS is given in Levshakov et al. (2003b), where the \HI\
column density of the DLA at \zabs =1.918 is also estimated. The Lyman
series of the \zabs = 2.917 LLS  allows  to establish a local continuum
shortward of $\lambda \sim 3575$ \AA\ and therefore to the damped \lya\
profile at \zabs = 1.918 (see upper panel of Fig.~1). This local continuum
has then been used to derive the \HI\ column density of the DLA system
of $\log N$(\HI) $\simeq 20.00$. This system is strictly speaking a sub-DLA
(P\'eroux et al. 2002), nevertheless, for completeness, we present the
results of the abundance determinations of metals associated with this
absorber.

In this system at \zabs = 1.9184, the \NI\ 1134 \AA\ multiplet falls near the
atmospheric cutoff and the \NI\ 1200 \AA\ multiplet is heavily blended
with \lya\ interlopers. Therefore the column density determination of
nitrogen is not possible.

\OI\ and \SiII\ column densities are constrained by 11 components whose
$z$ and $b$  have been priorly determined from the joint fit of
four \FeII\ unsaturated lines (Fig.~6). \OI\ 1302 \AA\ 
is saturated and we only derive an upper limit: 
$\log N$(\OI) $>$ 16.89,
by using the redshift and $b$-parameters of the 11 components obtained
from the \FeII\ transitions. The same procedure is applied to the
saturated \SiII\ 1260 \AA\ transition since the weaker \SiII\ 1808 \AA\
transition is contaminated on its right wing, and we find
$\log N$(\SiII) $>$ 15.42.  \SII\ transitions ($\lambda\lambda$ 1250,
1253, 1259 \AA) are all heavily blended with numerous \lya\ interlopers.
Several \FeII\ unsaturated lines are detected and have been used for
the column density determination: $\lambda\lambda$ 2344, 2382, 2586
and 2600 \AA. The resulting column density is 
$\log N$(\FeII) = 14.44$\pm$0.02. 
The \ZnII\ 2026, 2062 \AA\ transitions are not
detected and from the stronger $\lambda$ 2026 transition we estimate
for each component a 3$\sigma$ limit to the column density of 
$\log N$(\ZnII) $<$ 11.44. Table~7 provides a summary of the
fitting parameters and column density results.

The metallicity of this system as measured by iron is 
[Fe/H] = --1.06$\pm$0.20.  
Furthermore the $\alpha$ over Fe-peak ratio is high: 
we find [O/Fe] $>$ 1.21 and [Si/Fe] $>$ 0.92. 
Dust depletion  
may enhance the ratios involving iron, 
but not to the extent at which they are 
observed here  (see Vladilo 2002a).  
Considering the low \HI\ column density of the system
($\log N$(\HI) = 20.00$\pm$0.20), 
it is possible that the observed abundances are
also affected by ionization corrections. If that is the case, the true [O/Fe]
is expected to be lower than the observed one (e.g., Dessauges-Zavadsky et
al. {(2003)}. This could therefore explain our observational result. On the
contrary, the true [Si/Fe] ratio, is expected to be higher than the one
observed in the case where ionization is important. 
Nevertheless, it can be seen
from Fig.~6, that the \SiII\ profile is likely contaminated by
interlopers. Indeed, several absorptions to the blue and the red of the
profile are observed. Additional contaminations inside the profile itself
can not be excluded, although it looks saturated at first sight.

\subsection{DLA at \zabs=2.3377 towards QSO 1232+0815}

This DLA has been previously studied  by Srianand et al. (2000)
with the same UVES data, but they did not 
analyse the column densities of \NI\ and \SII\ which are measured
here for the first time.

Our derived neutral hydrogen column density, 
$\log N$(\HI) = 20.80$\pm$0.10 
obtained from the  \lya\ 
absorption is slightly lower than 
$\log N$(\HI) = 20.90$\pm$0.10 given by Srianand et al.
(2000).

The metal profiles are well fitted by 6 components, three of them 
composing the main absorption around  $v$ = 0  \kms (relative to
$z=2.3377$),
while the remaining three
components are weaker and only observed in 
the strongest metal absorptions towards the blue,  
around  $v$ = --60  \kms\  
(see Fig.~7). 
The simultaneous fit to 
the \FeII\  and \SiII\ absorptions, shown in the right panel of Fig.~7, 
is used
to fix the redshift $z$ and $b$-parameters of the six components. 
The derived $z$ and $b$-values are further used  to fit the 
\NI\ and \SII\ absorptions.  
The resulting parameters of the fit for each 
component are given in Table~8 along with the total column 
densities and abundances.

The total column densities of   \SiII\ and \FeII\ 
  are in agreement, within the errors, 
with those obtained by Srianand et al. (2000). 
For nitrogen we determine a total column density 
$\log N$(\NI) = 14.63$\pm$0.08 
only taking into account the 
components around  $v$ = 0  \kms, 
since the components at  $v \simeq$  --60 \kms\
are found to contribute at maximum 2\% of the total column density 
(well inside the  uncertainty interval of  the column density).
It is   the first time that the \SII\ column density is measured 
by means of high resolution data, and we obtain 
$\log N$(\SII) = 14.83$\pm$0.10
which is significantly larger (by 0.35 dex) than the value obtained by  
Ge et al. (2001) from low resolution spectra 
 (FWHM $\sim$ 60 \kms).  
In this  work they measure the \ZnII\ column density   and determine 
$\log N$(\ZnII) = 12.71$\pm$0.14 
which  we adopt  with some caution,   
since 
the \ZnII\ absorptions of this DLA fall outside the wavelength coverage
of the UVES spectra analysed  here. Using  this  \ZnII\ column density and 
our  \SII\ column density,  
we obtain [S/Zn] = --0.41$\pm$0.17
which would be the lowest value ever measured in a DLA. It is worth noting that
iron (from our analysis of the UVES data) is significantly underabundant 
with respect to Zn, [Fe/Zn] = --0.83, 
which would in turn indicates that iron is significantly  
depleted on dust. On the other hand, from our analysis of the UVES data we obtain a solar 
[Si/S] = --0.01 abundance ratio. If  dust depletion is severe 
in this DLA, as indicated by the [Fe/Zn] abundance ratio, then
we should expect an underabundance of Si with respect to S which 
is not observed.
The apparent contradiction between these 
two results may reside in the 
\ZnII\ column density determination derived from low resolution spectra.
For instance, if the \ZnII\ 2026 \AA\ absorption analysed by
Ge et al. (2001) is contaminated by atmospheric features, then 
the derived \ZnII\ column density will be overestimated and 
consequently the [S/Zn] and [Fe/Zn]  ratios underestimated.
A new determination of \ZnII\ 
column density by means of high resolution data is 
required in order to assess the degree of dust depletion 
and the abundance ratios in this DLA.

\section{The N/$\alpha$ abundance ratio in DLAs}

In order to study the behaviour of the N/$\alpha$ abundance ratios in DLAs,
we have collected all abundance measurements obtained
from high resolution spectra. 
In addition to the 5 new systems investigated  
here, we have compiled the DLAs with 
nitrogen and $\alpha$-capture element measurements available
in the literature, from both HIRES/Keck and UVES/VLT.
Our compilation,  containing  33 systems,  is  listed in Table~8. 

In Fig.~8 (upper panel),
we plot the [N/$\alpha$] versus [$\alpha$/H] ratios. The figure contains 
29 DLAs, since 4 systems with poorly stringent upper limits have 
been omitted for clarity.
We consider O, S and Si as three possible 
estimators of $\alpha$ abundance
measurements in DLAs.
In Fig.~8, abundance ratios are indicated with empty diamonds 
for [N/O] (7 cases), filled diamonds for [N/S]  (16 cases) and
filled squares for [N/Si] (6 cases). 
Our new measurements are indicated with additional circles.

When possible, we use O as the preferred $\alpha$-capture element.
An important advantage of this element is that the
N/O and O/H ratios are unaffected by  ionization corrections
since they are derived from the observed  \NI/\OI\ and \OI/\HI\ ratios, i.e.
from ratios between column densities of neutral species which
are not altered by an intervening \HII\ region, if present. 
In addition, oxygen, similarly to nitrogen, 
is not affected by dust depletion.  
The drawback is the difficulty in measuring 
accurate \OI\ column densities from unsaturated lines, as we mentioned in Section 3.  
Unsaturated \OI\ transitions lie inside the Ly$\alpha$ forest,
where it is necessary to detect at least two of them in order 
to ascertain that 
they are not Ly$\alpha$ interlopers.

The 5 N/O measurements, shown in  Fig.~8,  have been
derived using accurate column densities of \OI\, 
obtained from unsaturated or unblended transitions.  
Two stringent upper limits are also shown (see Table~8 for references). 

When reliable O data are not available,
we use  sulphur or, alternatively, silicon  
as $\alpha$-element indicators.  
Both are produced in the same
massive stars which produce oxygen and they
track each other over a wide range of metallicities in Galactic stars 
(Chen et al. 2002; Nissen et al. 2002, and
refs. therein)  
and in dwarf galaxies, including blue compact galaxies 
(Garnettt 1989; Skillman \& Kenicutt 1993; Skillman et al. 1994; 
Izotov \& Thuan 1999).  
One potential problem with S and Si in DLAs is that their abundances
are obtained from  \SII\ and \SiII\ ions 
which are dominant
ionization states in  \HI\ clouds, but can also arise in 
intervening \HII\ gas.

Izotov et al. (2001)  argue  
that intervening \HII\ gas is inherent to the DLAs ionization structure,  
yielding an increase of the
\SII\ and \SiII\ column densities without affecting the \NI\ 
column densities. As a consequence, the low
[N/S,Si] ratios observed in some DLAs could be an artifact of ionization 
effects rather than a genuine nucleosynthetic effect. 
However, the model proposed by Izotov et al. (2001) predicts an overabundance
of [Si/O] which is not observed in DLAs and this gives a strong evidence
against the general presence of intervening \HII\ regions 
(Vladilo et al. 2003). In fact, relative abundances of low ionization 
species in DLAs, including those
of \ion{Al}{iii}, can be explained
without invoking intervening \ion{H}{ii} regions, in which case 
ionization correction for \SII\ and \SiII\ 
are in general negligible (Vladilo et al. 2001).  
As noted by Molaro {(2003)}, the existence of very low values of the 
[N/O] ratio --- 
for which negligible contribution 
from intervening \HII\ gas is expected ---
indicates that the low [N/Si,S] ratios 
are not due to ionization effects. 
In fact, all the DLAs
with  low [N/O] have also   low [N/Si] values {(see Fig.~8 and 9)},
confirming that 
\SiII\ ionization corrections are not relevant.

In addition to ionization, dust depletion could 
alter the observed abundance ratios. Silicon
can be mildly depleted in DLAs  (Vladilo 2002b).  
Sulphur is not depleted on dust and for this reason is preferred 
to silicon as an indicator of $\alpha$-elements. 
If Si is depleted, the observed [N/Si] ratios are larger 
than the intrinsic ones.
Unfortunately, only one of the six DLAs of the [N/Si] sample has 
iron and zinc abundances, which are required to estimate the effect
of dust depletion. For this system, the DLA towards  QSO 1223+178,
we obtain a minimum possible intrinsic ratio of [N/Si] = --1.12 by
using  different dust-correction models 
%S00, S11, E00 and E11 
considered in Vladilo (2002a). 
The correction is inside the 1$\sigma$ error of the observed 
ratio [N/Si] = --1.01$\pm$0.18. 
By applying the same method to 
all DLAs with Si, Zn and Fe measurements,
we find that the Si/H depletion correction is generally small, with a mean value
$\simeq 0.1$ dex and even lower median value.
Therefore the Si depletion should not affect dramatically the [N/Si] plot.

In the bottom panel of  Fig.~8,  we show the data based  
on Si (20 measurements and 7 useful limits),
the $\alpha$-capture element most widely measured in DLAs.
By comparing the top and bottom panel of  Fig.~8  
(only 6 N/Si data points in common)
one can see that the use of Si instead of O and S  
does not significantly change the overall behaviour of the ratios, 
confirming that depletion effects for Si 
are not important for the majority of the DLAs shown in the figure. 

The nitrogen ratios in DLAs  are 
concentrated in two groups.
The majority of the [N/$\alpha$] values (22 out of 29) are distributed 
around {$\approx$ --0.9} dex, with typical metallicities
{[$\alpha$/H]  $\lsim$ --0.8} dex (hereafter `high-N DLAs'), while
the remaining 7 ratios have a relatively constant value around  
[N/$\alpha$] $\simeq -1.5$ dex and metallicities [$\alpha$/H] $\lsim -1.5$ dex
(hereafter `low-N DLAs'). 
The distribution of N/$\alpha$ ratios in two sub-samples was first suggested
by Prochaska et al. (2002), 
a result to be confirmed at that time, given the low number of
DLAs with low [N/$\alpha$] values in their sample 
(only two HIRES/Keck measurements). 
{One of our [N/$\alpha$] measurements (z=2.4 DLA in QSO 0841+129) shows an intermediate value
between  the low-N and high-N subsamples. However, as it can be seen in Fig. 9 
this DLA has a N abundance 2.5 times larger than the highest N abundance observed in the low-N group,
and for that reason we include it in the high-N subsample.}

We use the homogeneous sample of [N/Si] 
data to perform an analysis of the two groups. 
Considering only the 20 N/Si measurements, we find 
16 high-N DLAs, with a mean value  $<$[N/Si]$>$ = {--0.87 $\pm$ 0.16}   
(i.e., a standard deviation of {18\%} around the mean {($<$[N/Si]$>$ =  --0.84 $\pm$ 0.14, 
if we exclude the z=2.4 DLA in QSO 0841+129)}, 
and 4 low-N DLAs  with
$<$[N/Si]$> = -1.45\pm0.05$ (i.e., a standard deviation of 3\%).
Therefore, the two groups differ in their [N/Si] ratios by $\approx$ {0.6} 
dex on the average.
Both sub-samples are characterized by a low scatter of the [N/Si] ratios
and, in particular, the scatter of the low-N DLAs sub-sample is
essentially equal to the typical dispersion due to measurements errors. 
Even if the number of low-N DLAs is still small, it is 
worth noting that the  increase from 2 measurements plus 1 upper limit  
(in Prochaska et al. 2002) 
to 4 measurements plus 3 upper limits
(in the compilation presented here) still shows the 
[N/Si] ratios clustered around --1.5 dex
with very low dispersion,
confirming the existence of the low-N DLA sub-sample. 
{We stress that with the available instrumentation (Keck+HIRES and VLT2+UVES) one should 
have been able to detect 
\NI\ features at an abundance level [N/H] lower than that typical  of low-N DLAs.
For instance, 
the $3-\sigma$ detection limit is [N/H] $\simeq$ --3.9 
at the average \HI\ column density of the low-N DLAs  for a S/N $\simeq$ 25.
This [N/H] limit is nearly a factor 2 times lower
than the lowest value  observed so far (the \zabs=2.8 system in QSO\,1946+765).
Therefore the the lack of detection of DLAs at [N/H] $<$ 3.6 dex
probably reflects an intrinsic paucity of the population of
DLAs at very low values of [N/H].}

\section{Discussion}

To investigate the origin of N in DLAs  
we compare the observed abundance ratios involving this element with those
measured in extragalactic metal-poor \HII\ regions
and with predictions of galactic chemical evolution models.

Small dots in the
top panel of  Fig.~8  
indicate [N/O] vs. [O/H] measurements in metal-poor
\HII\ regions of dwarf galaxies 
which show a  plateau {with a mean value of [N/O] = --0.73 $\pm$ 0.13  
for metallicities [O/H] $\lsim$ --0.8, we used this value in correspondence with the 
matallicities of the DLAs with [N/Si] measurements}, 
(grey dots: Kobulnicky \& Skillman 1996; van Zee et al.
1996; van Zee et al. 1997; black dots: measurements in 
 BCD galaxies  by Izotov \& Thuan 1999).

The horizontal dashed line in  Fig.~8  
represents the average [N/O] ratio observed 
in  BCD galaxies  at low metallicities and is considered an  empirical
representation of the N/O level due to the primary production of N, 
since the N/O ratios are independent of metallicity. 
The tilted dashed line in the figure follows the [N/O] behaviour  
observed at [O/H] $\gsim$ --0.5 in dwarves and spiral galaxies
extrapolated to lower metallicities. This tilted line, 
is an empirical representation of the [N/O] rise due to the  
secondary production of N. 

In our sample the high-N DLAs (22 out of 29 systems) are distributed
around [N/$\alpha$] $\simeq$ {--0.9}, {comparable} to the [N/$\alpha$]  
plateau of  BCD galaxies, 
although DLAs extend to lower metallicities. 
Also the [N/$\alpha$] scatter in DLAs is 
comparable with the one observed in  BCD galaxies.    
On the other hand, the low-N DLAs have  
[N/$\alpha$] $\simeq -1.5$,
a factor of 5 lower than those measured in  BCD galaxies 
and, in fact, the lowest
ever observed in any astrophysical site.

\subsection{The high-N subsample of DLAs}

In the framework of chemical evolution models the plateau at
[N/O] $\simeq$ {--0.7/--0.9}   
is considered  as  an evidence of the primary production of N at low 
metallicities, since the N abundance increases in lockstep
with that of the primary oxygen. 
Most chemical evolution models 
 (Pilyugin 1999; Henry et al. 2000; 
Maynet \& Maeder 2002; Chiappini et al. 2002; Calura et al. 2002)  
supports the idea that the intermediate mass stars  
($4 < M/M_{\sun} < 8$)   
are the dominant sites of this primary production of N
(see, however, Izotov et al. 1999, 2001).
In these models
galaxies with continuous, 
low star formation rates  
(Henry et al. 2000)
as well as galaxies with star formation occuring 
in bursts separated by quiescent periods
(Pilyugin 1999)
can both reproduce the N/$\alpha$ ratios
observed in high-N DLAs and  BCD galaxies.  
In galaxies with low  star formation rate (SFR),  
the O and N abundance can increase in lockstep, if
the time required to achieve the low 
%O/H
metallicity observed is comparable to the  
lag time of intermediate mass stars to eject N. 
In galaxies with bursts,
the intermediate mass stars have enough time to 
deliver primary N in the same proportion
as O delivered by massive stars, 
if the quiescent periods between bursts are significantly
longer than the  
lag time of the intermediate mass stars 
for the ejection of N.

In the framework of these models, the high-N  plateau  
can be interpreted 
as an indication that the intermediate mass stars have already ejected the 
primary nitrogen synthesized by them.  
Even if the above models can reproduce the 
[N/$\alpha$] $\simeq$ {--0.9}  plateau observed 
in DLAs, they predict 
a very steep increase of the N/$\alpha$ ratios, 
 passing through 
the low-N DLAs ratios, but do not give a plateau 
at [N/$\alpha$] $\simeq$ --1.5.

\subsection{The low-N subsample of DLAs}

In previous  works on N abundances in DLAs where few measurements 
of low N/$\alpha$ ratios 
were available,
(Pettini et al. 1995; Lu et al. 1998; Centuri\'on et al. 1998;
Pettini et al. 2002),   
the range between the few low N/$\alpha$ ratios and several 
high N/$\alpha$ values was considered  to be a scatter   
due to the time-delay between the ejection of O by massive stars
and the one of primay N by intermediate mass stars.
The compilation of DLAs data presented here supports the  
idea of the existence of a sub-sample of DLAs
with [N/$\alpha$] values clustered around --1.5 dex with a very low scatter,
as suggested by Prochaska et al. (2002). 
As discussed in  Sect.~4, 
we are now confident that 
the low N/$\alpha$ ratios in DLAs are 
real and not due to ionization effects. 

These very low N/$\alpha$ ratios, only observed in DLAs, are a crucial 
evidence against the argument of Izotov et al. (1999, 2001), who 
proposed the primary production of N by massive stars to explain the 
plateau [N/$\alpha$] $\simeq$ {--0.7/--0.9} dex observed in 
 BCD galaxies  and the majority of DLAs.
In the case of primary N from massive stars,
the measured N/O values 
must constitute a lower envelope to any N/O ratio
observed in galaxies, 
because the intermediate mass stars 
can only increase the N/O at a later time 
(Pilyugin 1999; Pilyugin et al. 2002).
The present data suggest that 
the observational lower envelope is at [N/$\alpha$] $\simeq -1.5$,
the values found in low-N DLAs. 
Therefore, the primary production by massive stars  
cannot explain the values  [N/$\alpha$] $\simeq$ {--0.7/--0.9} dex
measured in  BCD galaxies  and high-N DLAs, favouring the idea that 
this plateau is due to the primary production of N by intermediate 
mass stars.   

In order to reproduce the low [N/$\alpha$] $\simeq -1.5$ values 
for these DLAs,
Prochaska et al. (2002)  invoked a truncated 
 initial mass function (IMF)  
at the low mass end 
 ($M \gsim 7M_{\sun}$).  
With such IMF, the galaxies would experience an initial burst 
of star formation where only massive stars are formed.
Chiappini et al. (2002)
have pointed out  that  a top-heavy IMF as proposed by 
Prochaska et al. (2002) should cause a strong  enhancement
of the $\alpha$/Fe-peak ratios in these low-N DLAs, 
which  in general is not observed
(see below).

\subsection{The [N/$\alpha$] values versus nitrogen abundances}

From the classical plot [N/$\alpha$] versus [$\alpha$/H] (see  Fig.~8 ), 
one can see that
low and high values of [N/$\alpha$] co-exist at a given  [$\alpha$/H].
This result, if confirmed by further data, apparently suggests the presence
of a bimodal distribution of [N/$\alpha$]  
ratios in DLAs, as proposed by Prochaska et al. (2002).
As shown by Molaro {(2003)},
the separation between  the two groups of DLAs 
appears more clearly by plotting [N/Si] versus the nitrogen abundance, [N/H]. 
This can be seen in Fig.~9, where we plot
[N/$\alpha$] (top panel) and [N/Si] (bottom panel) versus [N/H]
for the full, combined sample. 
Systems with nitrogen abundances [N/H] $\lsim -2.8$ 
appears to be separated from
those with [N/H] $\gsim -2.8$.  
Thus there is no  
overlap
between low and high  [N/$\alpha$] ratios at a given [N/H], suggesting that
the transition of the  [N/$\alpha$] ratios  appears  
at a particular value of the N abundance.  
To our knowledge,
the separation of DLAs in two groups
is not seen in other abundance measurements in these absorbers.
The fact that the division between low- and high-N DLAs 
appears at a particular value of nitrogen abundance  
suggests that the separation may result
from some peculiarity of the nitrogen enrichment history. 
 
The very low ratios ([N/$\alpha$] $\simeq -1.5$) 
observed in DLAs, together with their small scatter 
($\pm0.05$ dex), have led  Molaro {(2003)} to propose that this lower plateau
could be due to primary production of nitrogen in massive stars. 
If massive stars produce primary nitrogen, no time 
delay is expected between the injection of nitrogen  
and that of oxygen and,
as a consequence, a plateau with a small scatter of 
the N/O ratios is predicted (Pilyugin 1999).
The fact that the [N/$\alpha$] $\simeq -1.5$ values are the 
lowest ever observed
suggests that they may indeed represent the first observational evidence of 
primary production by massive stars. 
The low dispersion
of the  DLAs [N/$\alpha$] $\simeq -1.5$ values favours this 
interpretation, even if the present number of 
low-N measurements is still insufficient to establish firm conclusions.   
Clearly if other systems with even lower values of N/$\alpha$ are observed
the source of N in the systems at [N/$\alpha$] $\simeq -1.5$ will be called
into question.

As mentioned in Sect.~1,  the issue  
of primary production of nitrogen by massive stars is not 
settled. However, it is worth noting that the stellar models of 
Maynet \& Maeder (2002), 
predict N/O and O/H ratios in agreement with those  observed in  
the low-N DLAs by using a simple closed-box model and taking  
the integrated yields of massive stars (between 8 and 120 $M_{\sun}$)
with high rotational velocity  over a Salpeter IMF. 
Stellar evolution models of Umeda et al. (2000) for massive, 
metal-free  Population   III stars,
show that primary production of N may be significant. 
Their nitrogen yields as a function of metallicity give a 
[N/Mg] $\simeq -1.5$ for a 
$15M_{\sun}$ star
with  zero metallicity, or for a $20M_{\sun}$ star with a metallicity of
$Z/Z_{\sun} \simeq 0.05$.  
Even if these results from the literature do not provide
a conclusive evidence that massive stars are
 responsible for    
a primary production of N in the low-N DLAs,
at least they indicate 
that  primary N production in massive stars
is possible at the low level observed in the low-N DLAs, 
[N/O] $\simeq -1.5$. 

If primary nitrogen from massive stars is responsible for
the low values  [N/$\alpha$] $\simeq -1.5$  
measured in DLAs,  
then the ``classic" primary nitrogen plateau at  
[N/$\alpha$]$\simeq$ {--0.7/--0.9}  
must be due to intermediate 
mass stars, as claimed by the most recent works on chemical evolution. 
However, in these works the primary  production in intermediate mass stars 
cannot explain the very 
low values, [N/$\alpha$] $\simeq -1.5$,   observed in DLAs.

If future   estimates 
of the yields of primary N in massive stars will be able  to
explain the observed low ratios,
it will   not be necessary to invoke a top-heavy
IMF to explain the low-N DLAs. 
In this case, as Fig.~9  suggests, we might be seeing two different 
phases of the N enrichment in galaxies, rather than  galaxies
with a different
way of creating stars.
DLAs with [N/$\alpha$] $\simeq$ --1.5 
would be very young objects,  
caught before the ejection of primary N by intermediate mass stars,  
while DLAs with [N/$\alpha$] $\simeq$ {--0.9} will be older ones, caught 
after the lag time of the intermediate mass stars 
for the ejection of N. 
The {transition} between the low-N and high-N DLAs
could be linked to the short lag time ($\simeq$ 250 Myr,  see, e.g.,
Henry et al. 2000) of the  N enrichment by  intermediate mass stars. 
    
On the other hand, we cannot completely rule out  the possibility 
that {the lack of DLAs}  between the low-N and high-N {values observed in Fig. 9}, 
results from the 
limited statistics of the sample.
In this case we might be seeing 
the steep increase of N abundance in the course of evolution,  due to the primary N production in 
intermediate mass stars as most of the current models predict. 
If new measurements in DLAs at very low N abundance will yield ratios lower
than [N/$\alpha$] $\simeq$ --1.5 or  will fill the {separation}
between low- and high-N values, this would be a crucial evidence in
favour of this interpretation.
The problem with this scenario is the very short time scale ($< 250$ Myrs)
predicted for the increase of the [N/$\alpha$] ratios well above --1.5 dex (see for instance Fig. 3b in Henry et al. 2000),
which would make hard to detect low-N DLAs contrary to what observed.

\subsection{[N/$\alpha$] versus [$\alpha$/Fe] ratios}

The $\alpha$/Fe ratios is a classical indicator
of the galactic chemical evolution. The comparison of 
N/$\alpha$ versus $\alpha$/Fe ratios can be used to probe models of nitrogen
enrichment.
Qualitatively, if primary production of N
in massive stars is invoked to explain the very low N/$\alpha$
ratios, we would expect that these 
DLAs show an enhancement of the $\alpha$-elements, 
--- produced in the short-lived massive stars ending their lives 
as type-II supernovae  (SNe)  ---
relative to Fe-peak elements ---
ejected in longer time scales mainly by type-Ia SNe. 
It is worth keeping in mind that solar $\alpha$/Fe ratios are expected 
at low metallicities (comparable with those observed in DLAs)
when star formation proceeds in bursts  with long quiescent 
periods, or when star formation rates are low.
In both cases 
the metal enrichment is so slow that SNe Ia have enough time to evolve  
and enrich the medium 
with iron-peak elements balancing the $\alpha$-elements previously 
produced by SNe II, when the
overall metallicity is still low. 
If DLAs had a chemical evolution of this type, 
we  could expect an 
enhancement lower than [Si/Fe] $\simeq$ [S/Fe] $\simeq$ +0.3, the value  
observed at [Fe/H] $\simeq$ --1 (typically metallicity of DLAs 
as measured by Zn) 
in stars of our Galaxy (Chen et al. 2002), 
where the star formation has been much faster.

In analysing the $\alpha$/Fe ratios in DLAs care 
must be taken for dust depletion
effects, which may alter the Fe abundance and the $\alpha$/Fe ratio.
Unfortunately, none of the low-N DLAs have abundance determinations of   
Zn, the iron-peak element free from dust depletion. 
In Fig.~10, we plot [N/Si] versus [Si/Fe] for 
6 of the low-N DLAs\footnote{The remaining low-N DLA towards QXO0001 
 has  a poorly stringent upper limit on Fe abundance and hence 
[$\alpha$/Fe] abundance ratio (see Table 9).} 
({\it diamonds}) and also  the [N/Si] versus [Si/Zn]
ratios available for the high-N DLAs ({\it squares}). At first glance, 
the low-N DLAs show larger 
$\alpha$/Fe-peak ratios than the high-N DLAs.
The observed enhancement,
[Si/Fe] $\simeq +0.2/+0.4$, could  include
some contribution due to dust depletion. 
However, the low-N DLAs are all characterized by
very low metallicities, typically [Si/H] $< -1.5$, 
in which case dust effects are expected to be less critical. 
Therefore, some of the enhancement of the [Si/Fe] ratios
in low-N DLAs could be real. 
However, firm conclusions about the $\alpha$-enhancement cannot be drawn
with the present data, and measurements 
of Zn abundance in low-N DLAs are required in order to clarify this issue.

\section{Summary and conclusions}

The Nitrogen abundances in DLAs investigated here show the following characteristics:

1. The DLAs can be tentatively divided in two groups which differ in 
the [N/$\alpha$] ratios  by about {0.6} dex.

2. The high-N sub-sample contains  75\% of 
DLAs which show {a mean value of} [N/$\alpha$] $\simeq$ {--0.87}
with relatively low scatter ({0.16} dex).
The remaining DLAs belong to the low-N sub-sample, which     
shows  ratios clustered at {a mean value of} 
[N/$\alpha$] $\simeq$ --1.45 with an even lower dispersion (0.05 dex). 
The small number of low-N DLAs does not allow us to
conclude whether the [N/$\alpha$] ratios at --1.5 dex constitute
a plateau or not.

3. The low N/$\alpha$ values are not an effect of ionization.
 Direct measurements of \OI\ and \NI\ in DLAs for 
which negligible contribution 
from intervening \HII\ gas is expected, give low values of [N/O]
$\simeq$ --1.5, indicating that 
low [N/O] ratios are real. 
Moreover if ionization effects were relevant the  [Si/O] ratios obtained 
from \SiII\ and \OI\  
lines should be enhanced and this is not
observed in DLAs.

4. The low [N/$\alpha$] $\simeq -1.5$ observed in DLAs is the lowest 
value ever observed 
in any astrophysical site.  This is   
a crucial evidence against  the
primary production of  N by massive
stars, as  responsible for the plateau 
at [N/$\alpha$]$\simeq$ {--0.7/--0.9}  observed in 
 BCD galaxies  and in the majority of DLAs.  
 
5. The origin of the two groups of DLAs appears to be related to their N 
abundance and therefore
linked to the nucleosynthesis and enrichment history of this element.
The transition between the  two sub-samples occurs at [N/H] $\sim$ --2.8

6. Current standard models of chemical evolution can reproduce 
the [N/$\alpha$] $\simeq$ {--0.7/--0.9} 
plateau observed in BCD galaxies and high-N DLAs.  
These models pass through the low [N/$\alpha$] 
ratios, but do not give a  
plateau  at --1.5 dex. If further measurements in DLAs   will
give  more  [N/$\alpha$] values clustered at --1.5 dex,
the models will need      
to take into account this feature. 

7. A top heavy IMF has been invoked to explain the observed 
low-N subsample of DLAs.
We argue that the [N/$\alpha$] $\simeq$ --1.5 ratios and their 
low dispersion may suggest
a (modest) production of primary N in massive stars on top of which   
the primary production of N by intermediate mass stars {is seen}  which rises
the ratio to the value {--0.9} dex observed in the majority of DLAs. 
{In this scenario low-N DLAs 
would be very young objects,  
caught before the ejection of primary N by intermediate mass stars,  
while high-N DLAs will be older ones, caught 
after the lag time of the intermediate mass stars 
for the ejection of N.}

Nitrogen in DLAs  provides
unique indications on the earlier stages of galactic
evolution.  
{The} primary production of N deserves further 
investigation. Further constraints on our ideas of chemical evolution and
origin of nitrogen could be possible thanks to a full set of 
abundance measurements of nitrogen, 
$\alpha$-elements and Zn in individual DLA systems. In particular, observations 
towards higher redshift
systems will allow us to observe more DLAs at earlier stages of their 
evolution,  probably  
belonging to the low-N DLA class.

\begin{acknowledgements}
Special thanks are due to P. Bonifacio for making valuable comments 
on the manuscript.
We have benefitted from useful discussions with F. Calura and
C. Chiappini. 
CP is supported by a Marie Curie Fellowship. 
SAL is supported in part by the
RFBR grant No.~00-02-16007.
\end{acknowledgements}

\clearpage

%%%%%%%%%%%%%%%  TABLES %%%%%%%
%%%%%%%%%%%%%%%  TABLE2 %%%%%%%

\begin{table}
\caption{Summary {of} the atomic data used in the present analysis.}
\begin{tabular}{lr@{.}lr@{.}lc}
\hline
\hline
Transition &\multicolumn{2}{c}{$\lambda^a_{vac}$, \AA} & 
\multicolumn{2}{c}{f$_{\lambda}$} & Ref$^b$ \\ 
\hline
Ly-$\beta$  &  1025&7223 &  0&07912  & 1 \\
Ly-$\alpha$ &  1215&6701 &  0&41640  & 1 \\
NI 1134.1   &  1134&1653 &  0&01342  & 1 \\
NI 1134.4   &  1134&4149 &  0&02683  & 1 \\
NI 1134.9   &  1134&9803 &  0&04023  & 1 \\
NI 1199.5   &  1199&5496 &  0&13280  & 1 \\
NI 1200.2   &  1200&2233 &  0&08849  & 1 \\
NI 1200.7   &  1200&7098 &  0&04423  & 1 \\
NI 953.4    &   953&4152 &  0&01063  & 1 \\
OI 1025     &  1025&7616 &  0&01700  & 1 \\
OI 1026     &  1026&4757 &  0&00246  & 1 \\
OI 1039     &  1039&2303 &  0&00920  & 1 \\
OI 1302     &  1302&1685 &  0&04887  & 1 \\
OI 1355     &  1355&5977 &  1&248E-06& 1 \\
SiII 989    &   989&8731 &  0&13300  & 1 \\
SiII 1020   &  1020&6989 &  0&02828  & 1 \\
SiII 1190   &  1190&4158 &  0&25020  & 1 \\
SiII 1193   &  1193&2897 &  0&49910  & 1 \\
SiII 1260   &  1260&4221 &  1&00700  & 1 \\
SiII 1304   &  1304&3702 &  0&08600  & 2 \\
SiII 1526   &  1526&7065 &  0&11000  & 2 \\
SiII 1808   &  1808&0126 &  0&00218  & 3 \\
SII 947	    &   946&9780 &  0&00498  & 4 \\
SII 1250    &  1250&5840 &  0&00545  & 1 \\
SII 1253    &  1253&8110 &  0&01088  & 1 \\
SII 1259    &  1259&5190 &  0&01624  & 1 \\
FeII 1055   &  1055&2617 &  0&00800  & 1 \\
FeII 1062   &  1062&1520 &  0&00380  & 1 \\
FeII 1063   &  1063&1764 &  0&05998  & 1 \\
FeII 1063.9 &  1063&9720 &  0&00370  & 5 \\
FeII 1081   &  1081&8748 &  0&01400  & 1 \\
FeII 1083   &  1083&4204 &  0&00406  & 1 \\
FeII 1085   &  1085&1381 &  0&00015  & 1 \\
FeII 1096   &  1096&8770 &  0&03199  & 1 \\
FeII 1106.2 &  1106&2208 &  0&00001  & 1 \\
FeII 1106.3 &  1106&3596 &  0&00150  & 1 \\
FeII 1110   &  1110&2803 &  0&00112  & 1 \\
FeII 1112   &  1112&0480 &  0&00620  & 5 \\
FeII 1121   &  1121&9749 &  0&02000  & 1 \\
FeII 1125   &  1125&4478 &  0&01600  & 5 \\
FeII 1127   &  1127&0984 &  0&00300  & 1 \\
FeII 1133   &  1133&6650 &  0&00600  & 1 \\
FeII 1142   &  1142&3656 &  0&00420  & 5 \\
FeII 1143   &  1143&2260 &  0&01331  & 1 \\
FeII 1144   &  1144&9379 &  0&10500  & 1 \\
FeII 1260   &  1260&5330 &  0&02500  & 1 \\
FeII 1588   &  1588&6876 &  0&00012  & 1 \\
FeII 1608   &  1608&4510 &  0&05800  & 6 \\
FeII 1611   &  1611&2004 &  0&00130  & 6 \\
FeII 1901   &  1901&7729 &  0&00010  & 1 \\
FeII 2249   &  2249&8767 &  0&00182  & 7 \\
FeII 2260   &  2260&7805 &  0&00244  & 7 \\
FeII 2344   &  2344&2141 &  0&11400  & 6 \\
FeII 2374   &  2374&4612 &  0&03130  & 6 \\
FeII 2382   &  2382&7649 &  0&30060  & 1 \\
FeII 2586   &  2586&6499 &  0&06840  & 8\\
FeII 2600   &  2600&1729 &  0&22390  & 1 \\
ZnII 2026   &  2026&1360 &  0&48860  & 9\\
ZnII 2062   &  2062&6641 &  0&25640  & 9\\
\hline
\end{tabular}
\\
\scriptsize{
$^a$Vacuum rest wavelength \\
$^b$REFERENCES --- 1: Morton (1991); 2: Spitzer \& Fitzpatrick (1993); 3:
Bergeson \& Lawler (1993b); 4: Bonifacio et al. (2001);   
5: Howk et al. (2000); 6: Welty et al. (1999); 7: 
Bergeson et al. (1994); 8: Cardelli \& Savage (1995); 9:
Bergeson \& Lawler (1993a). 
}
\end{table}

\clearpage

%%%%%%%%%%%%%%%  TABLE3 %%%%%%%

\begin{table}
\label{t:Q0528-z2.1}
\caption{Total column densities and metal
abundances in the \zabs = 2.1410 DLA
towards PKS 0528$-$250}
%\begin{center}
\begin{tabular}{llcc}
\hline
\hline
Ion & $\lambda_{rest}$, \AA &  $\log N$(X), cm$^{-2}$  & [X/H]     \\
\hline
\HI	&1215	&20.95$\pm$0.05	&	\\
\NI   	&1134.4 &14.58$\pm$0.08	& --2.30$\pm$0.09 	\\
      	&1134.9	\\
\SiII   &1304  	&15.22$\pm$0.05	& --1.29$\pm$0.07	\\
  	&1526  \\
  	&1808  \\
\SII    &1253  	&14.83$\pm$0.04	& --1.32$\pm$0.06	\\
\FeII   &1125  	&14.85$\pm$0.09 	& --1.60$\pm$0.10	\\
       	&1608\\
       	&1611\\
       	&2344\\
       	&2374\\
       	&2586\\
\ZnII   & 2062  &$\leq$ 12.13           & $\leq -1.01$  \\
\hline
\end{tabular}
%\\ \scriptsize{ $^a$ The column densities of \OI\ and \ZnII\ cannot be
%derived in this system. \\
%$^b$ The two last components are not present in \SII.\\
%}
\end{table}

%\clearpage

%%%%%%%%%%%%% TABLE 4

\begin{table}[t]
\label{t:Q0528-z2.8}
\caption{Total column densities and metal
abundances in the \zabs = 2.8120 DLA
towards PKS 0528$-$250}
%\begin{center}
\begin{tabular}{llcc}
\hline
\hline
Ion & $\lambda_{rest}$, \AA & $\log N$(X), cm$^{-2}$  & [X/H]  \\
\hline
\HI	&1215	&21.11$\pm$0.04 	&	\\
\NI	&953.4 	&$\leq$15.66		& $\leq$ --1.38	\\
	&1134.1	\\
	&1134.4	\\
	&1134.9	\\
	&1199.5	\\
	&1200.2	\\
	&1200.7	\\
\SiII   &1808   &16.01$\pm$0.03   	& --0.66$\pm$0.06     \\
\SII    &1250   &15.56$\pm$0.02	& --0.75$\pm$0.04	\\
	&1253	\\
\FeII   &1081	&15.47$\pm$0.02	& --1.14$\pm$0.04	\\
       	&1096	\\
	&1125	\\
	&1608	\\
	&2374	\\
\ZnII   &2026  	&13.27$\pm$0.03	& --0.51$\pm$0.05	\\
\hline
\\
\end{tabular}
%\\ \scriptsize{ $^a$ The column densities of \OI\ and \SiII\ cannot be
%derived in this system. \\
%
%$^b$ The last two components are not present in \NI.
%}
\end{table}

%\clearpage

%%%%%%%%%%%%%%%  TABLE5 %%%%%%%
%%%%%%%%%%%%  z=2.3745
\begin{table}[!h]
\caption{Total column densities and metal
abundances in the \zabs = 2.3745 DLA towards QSO 0841+129}
%\begin{center}
\begin{tabular}{llcc}
\hline
\hline
Ion & $\lambda_{rest}$, \AA & $\log N$(X), cm$^{-2}$ & [X/H] \\
\hline
\HI   & 1215    & 21.00$\pm$0.10 & \\
\NI   & 1134.4  & 14.62$\pm$0.03 &  --2.31$\pm$0.10\\
     &   1134.9\\
\SiII & 1304  & $\geq$ 14.65$^a$  & $\geq$ --1.91$^a$   \\
%\PII    &     1152  & 2.374539   & $\leq$ 12.70  &   & $\leq$ --1.86\\        
\SII & 1259   & 14.77$\pm$0.03 &  --1.43$\pm$0.10  \\
%\ArI &1048 &2.374591&13.46$\pm$0.10&12.9$\pm$2.6&--2.06$\pm$0.15\\
%\CrII&2062&2.374523&13.14$\pm$0.02&13.2$\pm$1.1&--1.55$\pm$0.10\\
\FeII & 1081  & 14.87$\pm$0.04 &  --1.63$\pm$0.11\\
       &  1121\\
       & 1125\\
       &  1133\\
       & 1143\\
       & 1144 \\
\ZnII  & 2026  & 12.20$\pm$0.05 & --1.47$\pm$0.11\\
\hline
\\
\end{tabular}
\\
\scriptsize{
%$^a$	OI $\lambda\lambda$ 1039,1355 \AA\ 
%are not detected while 1302 \AA\ is as usually 
%heavily saturated \\
%
$^a$Prochaska \& Wolfe (1999) obtained from the unsaturated 
\SiII\ 1808 \AA\ transition 
$\log N$(\SiII) = 15.24$\pm$0.03, which yields [Si/H] = --1.32$\pm$0.10.
}
\end{table}

%\clearpage

%%%%%%%%%%%  TABLE 6

\begin{table}[!h]
\caption{Total column densities and metal abundances 
in the \zabs = 2.4762 DLA towards QSO 0841+129}
%\begin{center}
\begin{tabular}{llcc}
\hline
\hline
Ion & $\lambda_{rest}$, \AA & $\log N$(X), cm$^{-2}$ &  [X/H] \\
\hline
\HI   & 1215     & 20.78$\pm$0.08      & \\
      & 1025\\
\NI   & 1134.4  &14.12$\pm$0.03 &  --2.59$\pm$0.09 \\
      & 1200.2\\
      & 1200.7\\
\SiII & 1020    & 14.95$\pm$0.10 &  --1.39$\pm$0.13\\
%\PII & 1152 & &$\leq$12.45 & & $<$ --1.94\\        
\SII  & 1250  & 14.59$\pm$0.03 &  --1.39$\pm$0.09\\
      & 1253  \\
      & 1259  \\
%\ArI & 1048&2.476176&13.34$\pm$0.12&12.0$\pm$1.5&--1.96 $\pm$ 0.16\\
%\CrII& 2062 & &$<$ 13.22$^b$  & &  $<$--1.25          \\
%$^b$ PW99  gave log\,N(\CrII) = 12.84 $\pm$0.04  which yields 
%[Cr/H] = --1.63 $\pm$0.11   \\ 
\FeII & 1081 & 14.55$\pm$0.07 &  --1.73$\pm$0.11\\
      & 1096\\
\ZnII & 2026 & $<$ 12.13$^a$  & $<$ --1.32 \\
\hline
\\
\end{tabular}
\\
\scriptsize{
%$^a$OI $\lambda\lambda$ 1039,1302 \AA\ are heavily saturated \\
%$^b$Central wavelength have been fixed at the redshift observed 
%in the \SII\ triplet \\ 
%
$^a$Prochaska \& Wolfe (1999) obtained a more restrictive limit 
$\log N$(\ZnII) $<$ 11.78, which yields 
[Zn/H] $<$ --1.67 \\ 
}
\end{table}

%\clearpage

%%%%%%%%%%%  TABLE 7

\begin{table}[h!]
\caption{Total column densities and mtal abundances 
in the \zabs = 1.9184 DLA towards HE 0940$-$1050}
\label{t:Q0940}
%\begin{center}
\begin{tabular}{llcc}
\hline
\hline
Ion & $\lambda_{rest}$, \AA & $\log N$(X), cm$^{-2}$ & [X/H]    \\
\hline
\HI   &1215 & 20.00$\pm$0.20 & \\
\OI   &1302 & $>$ 16.89      & $>$ 0.15 \\
\SiII &1260 & $>$ 15.42      & $>$ --0.14 \\
\FeII &2344 & 14.44$\pm$0.02 & --1.06$\pm$0.20\\
      &  2382\\
      & 2586\\
      & 2600\\
\ZnII & 2026 & $<$ 12.48     &  $<$ --0.19  \\
\hline
\\
\end{tabular}
%\\ \scriptsize{ $^a$ The column density of \NI\, and \SII\  
%cannot be derived in this system. 
%The transitions of these elements are heavily blended
%and a limit to the column density cannot be estimated.} 
\end{table}

%\clearpage

%%%%%%  TABLE 8  Q1232
  
\begin{table}[h!]
\caption{Total column densities and metal abundances in  
the \zabs = 2.3377 DLA towards QSO 1232+0815 }
%\begin{center} 
\label{t:q1232}
%\begin{center}
\begin{tabular}{llcc}
\hline
\hline
Ion & $\lambda_{rest}$, \AA & $\log N$(X), cm$^{-2}$ & [X/H]  \\
\hline
\HI    &1215    & 20.80$\pm$0.10  & \\
\NI    &1134.1  & 14.63$\pm$0.08  & --2.10$\pm$0.13   \\
       &1134.4 \\
       &1134.9 \\
       &1199.5 \\
       &1200.2 \\
       &1200.7 \\
\SiII  &1304     & 15.18$\pm$0.09  &  --1.18$\pm$0.13  \\
       &1526\\
\SII   &1250     & 14.83$\pm$0.10 &  --1.17$\pm$0.14   \\
       &1253 \\
       &1259 \\
\FeII  &1125     & 14.71$\pm$0.08 &  --1.59$\pm$0.13   \\
       &1608\\
       &1611\\
\hline
\\
\end{tabular}
\\ 
%\scriptsize{ 
%$^a$ NI 1134.1,1134.4, 1134.9,  and NI 1199.5,1200.2,1200.7  transitions\\ 
%$^b$ OI $\lambda\lambda$ 1039,1302 \AA\  are heavily saturated \\
%
%$^c$\ZnII\ transitions are not covered in the UVES spectra analysed here. 
%From low resolution spectra (FWHM $\simeq 50$ \kms), Ge et al. (2001) 
%obtain a total column
%density of 
%$\log N$(\ZnII) = 12.71$\pm$0.14, yielding [Zn/H] = -0.76$\pm$0.17.  
%}
\end{table}

\clearpage

%%%%%%%%%%
%%%	 SUMMARY TABLE  TABLE 9
%%%%%%%%%%

\begin{table*}[!h]
\caption{N and $\alpha$-elements abundance measurements in DLAs}
%%%\begin{center}
\begin{tabular}{llcccccccc}
 \hline
 \hline
QSO&\zabs&$\log N$(\HI),&$\log N$(\NI),&[N/$\alpha$]&[$\alpha$/H]&$\alpha$- & [N/Si] & [Si/H] & Ref$^b$  \\
   &   & cm$^{-2}$ &  cm$^{-2}$ &  &  & elem. \\ 
\hline
 0000--263  & 3.390 & 21.41 $\pm$ 0.08 & 14.73 $\pm$ 0.02 & --0.73 $\pm$ 0.05 & --1.88 $\pm$ 0.09  & O     & --0.70 $\pm$ 0.03 & --1.91 $\pm$ 0.08 & 1	  \\
 0100+130  & 2.309 & 21.37 $\pm$ 0.08 & 15.03 $\pm$ 0.02 & --0.81 $\pm$ 0.02 & --1.46 $\pm$ 0.08  & S	 & --0.93 $\pm$ 0.05 & --1.34 $\pm$ 0.09 & 2,3	  \\
 0201+1120 & 3.386 & 21.26 $\pm$ 0.06 & 15.33 $\pm$ 0.06 & --0.61 $\pm$ 0.09 & --1.25 $\pm$ 0.09  & S	 &                  &                  & 4	  \\
 0201+365  & 2.462 & 20.38 $\pm$ 0.05 & $>$ 15.00	  & $>$ --1.02        & --0.29 $\pm$ 0.05  & S	 & $>$ --0.90        & --0.41 $\pm$ 0.05 & 2	  \\
 0307--4945 & 4.466 & 20.67 $\pm$ 0.09 & 13.57 $\pm$ 0.12 & --1.53 $\pm$ 0.21 & --1.50 $\pm$ 0.19  & O	 & --1.48 $\pm$ 0.14 & --1.55 $\pm$ 0.11 & 5	  \\
 0336--0142 & 3.062 & 21.20 $\pm$ 0.10 & $>$ 15.04	  & $>$ --0.68        & --1.41 $\pm$ 0.10  & S	 &                  &                  & 2	  \\
 0347--383  & 3.025 & 20.63 $\pm$ 0.01 & 14.89 $\pm$ 0.01 & --0.94 $\pm$ 0.01 & --0.73 $\pm$ 0.01  & O	 & --0.72 $\pm$ 0.01 & --0.95 $\pm$ 0.01 & 6	  \\
 0528--250  & 2.141 & 20.95 $\pm$ 0.05 & 14.58 $\pm$ 0.08 & --0.98 $\pm$ 0.09 & --1.32 $\pm$ 0.06  & S	 & --1.01 $\pm$ 0.09 & --1.29 $\pm$ 0.07 & 7	  \\
 0528--250  & 2.811 & 21.11 $\pm$ 0.04 & $<$ 15.66	  & $<$ --0.63	     & --0.75 $\pm$ 0.04  & S	 & $<$ --0.71        & --0.66 $\pm$ 0.07 & 7	  \\
 0741+4741 & 3.017 & 20.48 $\pm$ 0.10 & 13.98 $\pm$ 0.01 & --0.75 $\pm$ 0.02 & --1.68 $\pm$ 0.10  & S	 & --0.74 $\pm$ 0.04 & --1.69 $\pm$ 0.11 & 2	  \\
 0841+129  & 2.374 & 21.00 $\pm$ 0.10 & 14.62 $\pm$ 0.03 & --0.88 $\pm$ 0.04 & --1.43 $\pm$ 0.10  & S	 & --0.99 $\pm$ 0.04 & --1.32 $\pm$ 0.10 & 7	  \\
 0841+129  & 2.476 & 20.78 $\pm$ 0.08 & 14.12 $\pm$ 0.03 & --1.20 $\pm$ 0.04 & --1.39 $\pm$ 0.09  & S	 & --1.20 $\pm$ 0.10 & --1.39 $\pm$ 0.13 & 7	  \\
 0930+2858 & 3.235 & 20.30 $\pm$ 0.10 & 13.74 $\pm$ 0.01 & --0.66 $\pm$ 0.15 & --1.83 $\pm$ 0.18  & S	 & --0.51 $\pm$ 0.02 & --1.98 $\pm$ 0.10 & 2	  \\
 1055+4611 & 3.317 & 20.34 $\pm$ 0.10 & $<$ 14.09	  & $<$ --0.57	     & $>$--1.61	         & Si$^a$& $<$ --0.57        & $>$ --1.61        & 3	  \\
 1117--1329 & 3.351 & 20.84 $\pm$ 0.12 & $<$ 14.53        & $<$ --0.97        & --1.27 $\pm$ 0.13  & Si    & $<$ --0.97        & --1.27 $\pm$ 0.13 & 8        \\
 1122--1648 & 0.681 & 20.45 $\pm$ 0.15 & $<$ 14.50	  & $<$ --1.23	     & --0.65 $\pm$ 0.19  & Si	 & $<$ --1.23        & --0.65 $\pm$ 0.19 & 9	  \\
 1202--0725 & 4.382 & 20.60 $\pm$ 0.05 & 13.80 $\pm$ 0.10 & --0.96 $\pm$ 0.07 & --1.77 $\pm$ 0.07  & Si	 & --0.96 $\pm$ 0.07 & --1.77 $\pm$0.07 & 10	  \\
 1223+178  & 2.465 & 21.50 $\pm$ 0.10 & 14.83 $\pm$ 0.18 & --1.01 $\pm$ 0.18 & --1.59 $\pm$ 0.10  & Si	 & --1.01 $\pm$ 0.18 & --1.59 $\pm$ 0.10 & 2,11	  \\
 1232+0815 & 2.337 & 20.80 $\pm$ 0.10 & 14.63 $\pm$ 0.08 & --0.93 $\pm$ 0.13 & --1.17 $\pm$ 0.14  & S	 & --0.92 $\pm$ 0.12 & --1.18 $\pm$ 0.13 & 7	  \\
 1331+170  & 1.776 & 21.18 $\pm$ 0.04 & 14.84 $\pm$ 0.10 & --1.01 $\pm$ 0.14 & --1.26 $\pm$ 0.11  & S	 & --0.82 $\pm$ 0.11 & --1.45 $\pm$ 0.06 & 2,11,12,13\\
 1409+0950 & 2.456 & 20.54 $\pm$ 0.10 & $<$ 13.19 	  &  $<$ --1.15       & --2.13 $\pm$ 0.12  & O	 & $<$ --1.26        & --2.02 $\pm$ 0.10 & 14	  \\
 1425+6039 & 2.826 & 20.30 $\pm$ 0.04 & 14.70 $\pm$ 0.01 & $<$ --0.50        & $>$ --1.03        & Si$^a$& {$<$ --0.50}        & $>$ --1.03        & 2\\
 1759+75   & 2.625 & 20.80 $\pm$ 0.01 & 14.99 $\pm$ 0.03 & --0.98 $\pm$ 0.03 & --0.76 $\pm$ 0.10  & S	 & --0.92 $\pm$ 0.03 & --0.82 $\pm$ 0.10 & 9,12,15 \\
 1946+765  & 2.844 & 20.27 $\pm$ 0.06 & 12.59 $\pm$ 0.04 & --1.42 $\pm$ 0.04 & --2.19 $\pm$ 0.06  & O	 & --1.38 $\pm$ 0.04 & --2.23 $\pm$ 0.06 & 2,11	  \\
 2059--360  & 2.507 & 20.21 $\pm$ 0.10 & 12.66 $\pm$ 0.18 & --1.44 $\pm$ 0.21 & --2.04 $\pm$ 0.15  & Si	 & --1.44 $\pm$ 0.21 & --2.04 $\pm$ 0.15 & 16	  \\
 2206--199  & 2.076 & 20.43 $\pm$ 0.06 & $<$ 12.88	  & $<$ --1.55        & --1.93 $\pm$ 0.11  & O	 & $<$ --1.14        & --2.34 $\pm$ 0.06 & 14	  \\
 2212--1626 & 3.662 & 20.20 $\pm$ 0.08 & $<$ 13.58	  & $<$ --0.70        & --1.85 $\pm$ 0.08  & Si$^a$& $<$ --0.70      & --1.85 $\pm$ 0.08 & 3,17	  \\
 2233+1310 & 3.149 & 20.00 $\pm$ 0.10 & $<$ 14.32	  & $<$ --0.56        & $>$ --1.05         & Si$^a$& {$<$ --0.56}  & $>$ --1.05        & 3	  \\
 2243--6031 & 2.330 & 20.67 $\pm$ 0.02 & 14.88 $\pm$ 0.10 & --0.87 $\pm$ 0.14 & --0.85 $\pm$ 0.10  & S	 & --0.85 $\pm$ 0.10 & --0.87 $\pm$ 0.03 & 18	  \\
 2343+1232 & 2.431 & 20.35 $\pm$ 0.05 & 14.63 $\pm$ 0.05 & --0.84 $\pm$ 0.07 & --0.81 $\pm$ 0.07  & S	 & --0.81 $\pm$ 0.07 & --0.84 $\pm$ 0.07 & 19,20	  \\
 2344+1228 & 2.537 & 20.36 $\pm$ 0.10 & 13.78 $\pm$ 0.03 & --0.77 $\pm$ 0.03 & --1.74 $\pm$ 0.10  & Si	 & --0.77 $\pm$ 0.03 & --1.74 $\pm$ 0.10 & 2	  \\
 2348--147  & 2.279 & 20.56 $\pm$ 0.08 & $<$ 13.22	  & $<$ --1.24        & --2.03 $\pm$ 0.14  & S	 & $<$ --1.35        & --1.92 $\pm$ 0.08 & 2	    \\
 QXO0001   & 3.000 & 20.70 $\pm$ 0.05 & 13.32 $\pm$ 0.04 & --1.64 $\pm$ 0.04 & --1.67 $\pm$ 0.05  & O	 & --1.50 $\pm$ 0.04 & --1.81 $\pm$ 0.05 & 2	   \\
\hline

\\
\end{tabular} 
 \\
 \scriptsize{
$^a$These values are {not} plotted in  Figures 8,9,10  
since limits are not restrictive enough \\
$^b$REFERENCES --- 1: Molaro et al. (2001); 2: Prochaska et al. (2002); 
3: Lu et al. (1998); 4 Ellison et al. (2001);
5: Dessauges-Zavadsky et al. (2001); 6: Levshakov et al. (2002); 
7: This work; 8: P\'eroux et al. (2002); 9: de la Varga et al. (2002);
10: D'Odorico et al. (2003);
11: Prochaska et al. (2001); 12: Prochaska \& Wolfe (1999); 
13: Kulkarni et al. (1996); 14: Pettini et al. (2002); 
15: Outram et al. (1999); 16: Dessauges-Zavadsky et al. {(2003)}; 
17: Lu et al. (1996); 18: Lopez et al. (2002);
19: D'Odorico et al. (2002); 20: D'Odorico V. (private communication), 
    the N, S, and Si column densities have been recomputed using 
    Morton (1991) oscillator strength values.  
}
\end{table*}

%%%%%%%%%%%%%%%%%%%%%%%%%%%%%%%%%%%%%
%%%%%%            FIGURES
%%%%%%%%%%%%%%%%%%%%%%%%%%%%%%%%%%%%%

%----------------Fig. 1
\clearpage

\vspace{2cm}
\begin{figure*}[!h]
\begin{center}
\includegraphics[clip=true,width=15.5cm,height=22cm]{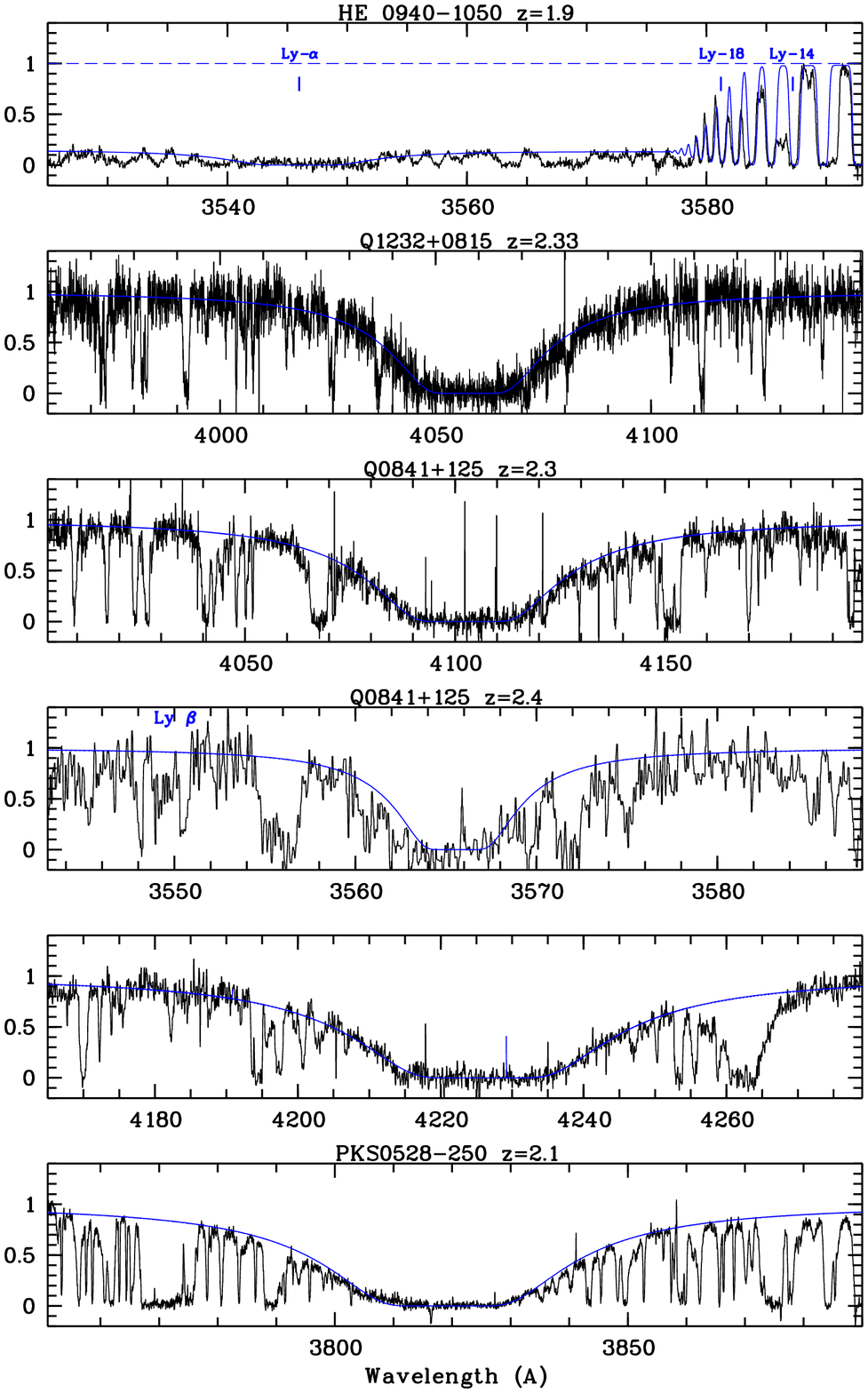}
\end{center}
\caption{\HI\ \lya\ absorptions in the normalized UVES spectra of the QSOs studied here.  
The top panel shows the \lya\ of the  \zabs = 1.918 DLA system 
towards HE 0940--1050 with local continuum
determined from a full fit to the Lyman series of the \zabs = 2.917 LLS (Levshakov et al. 2003b).
For the \zabs = 2.476 DLA system towards QSO 0841+129, 
the \lyb\ absorption is also shown.
Smooth lines are the synthetic spectra obtained from
the fit giving the $N$(\HI) column densities presented 
in Tables 3 to 7}
\end{figure*}

%-----------------------Fig. 2
\clearpage

\begin{figure*}[!h]
\begin{center}
\vspace{0.4cm}
\includegraphics[clip=true,angle=-90,width=11cm]{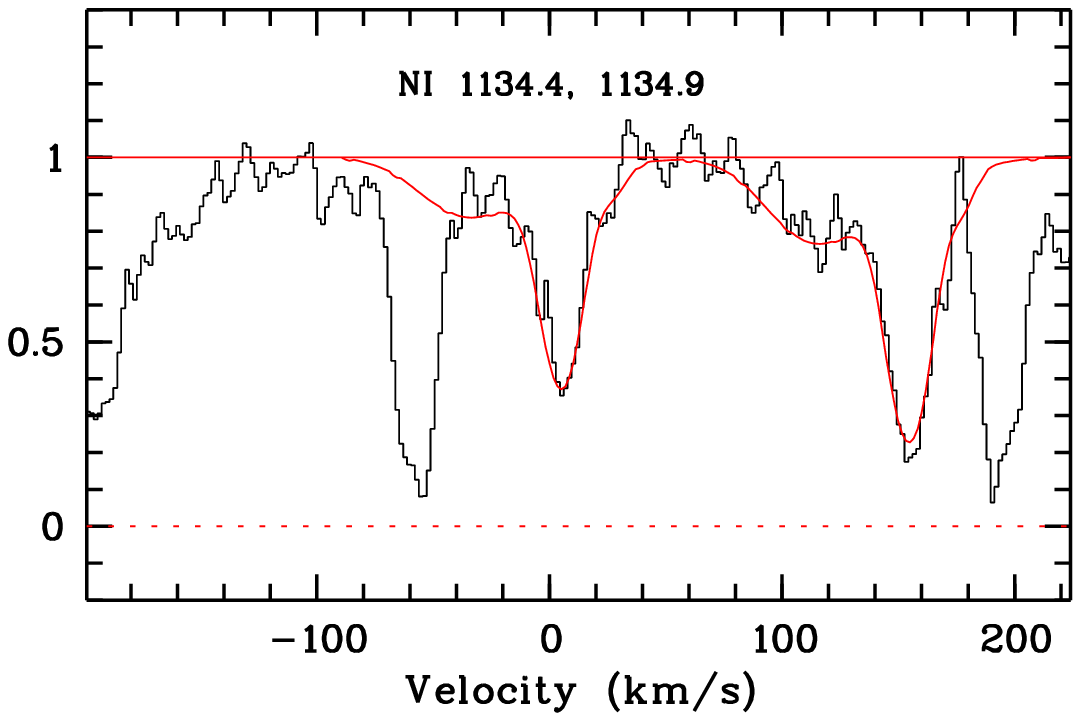}
\end{center}
\vspace{2cm}
\includegraphics[clip=true,angle=-90,width=8.2cm]{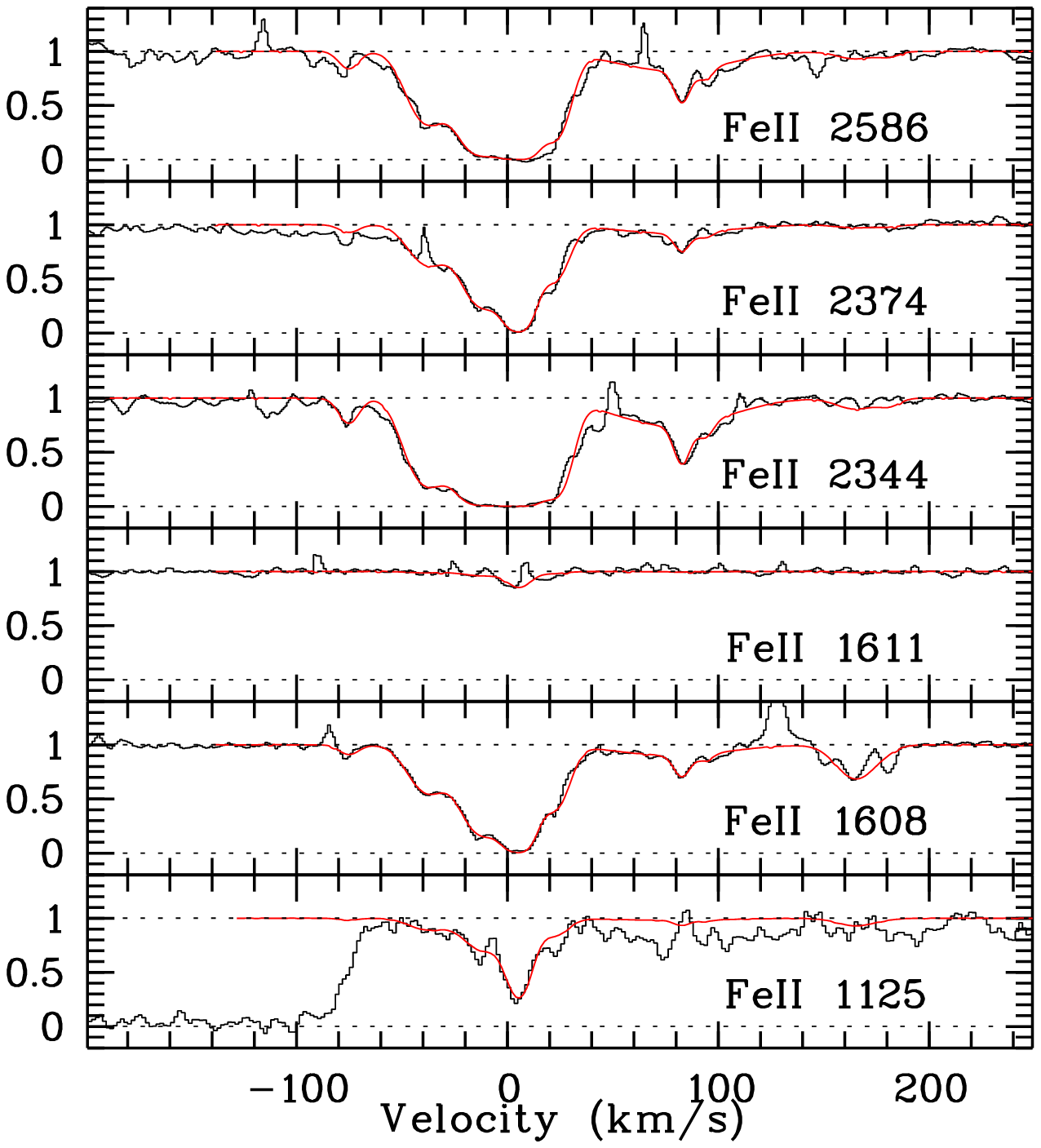}
\hspace{1.2cm}
\includegraphics[clip=true,angle=-90,width=8.2cm]{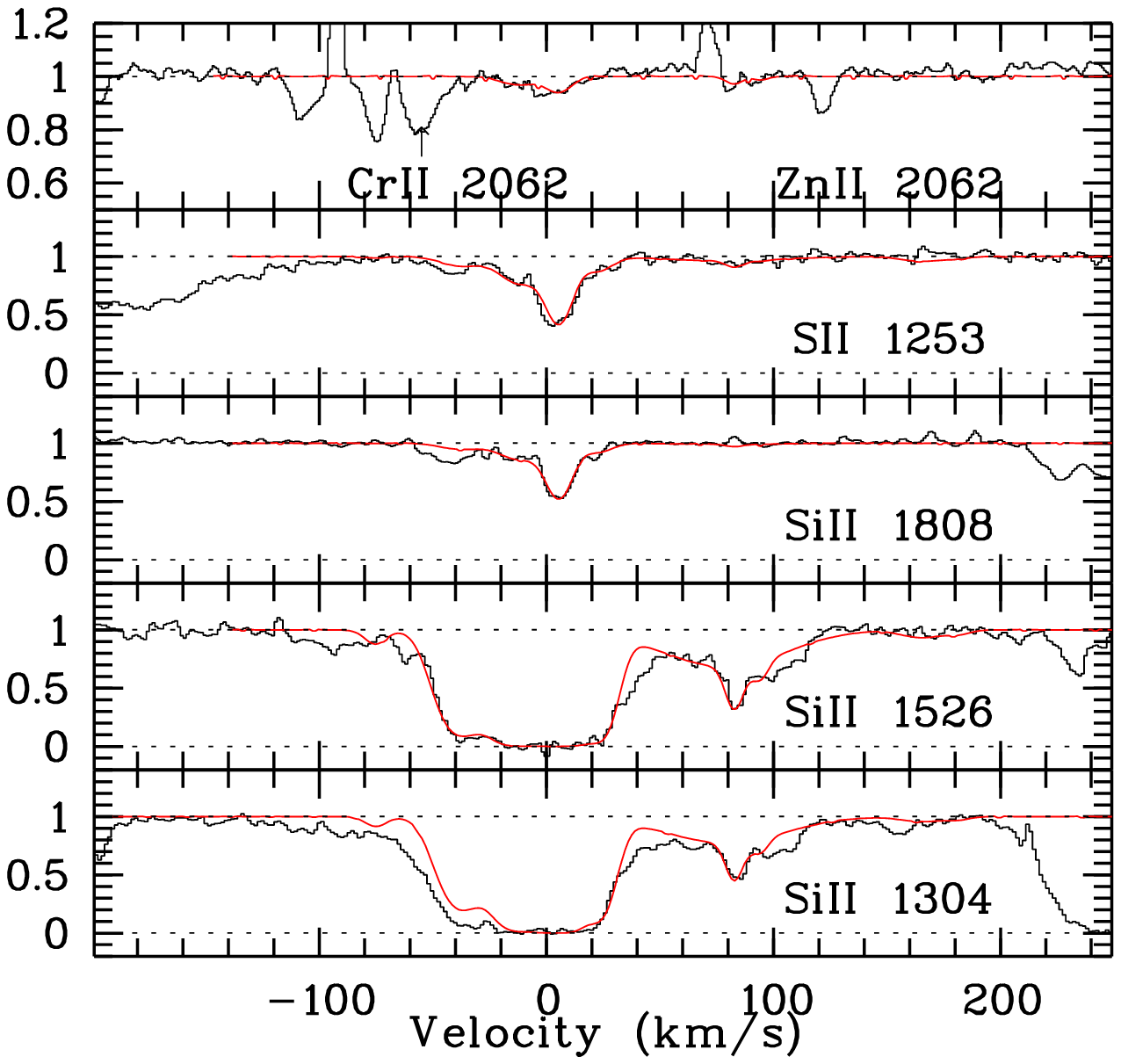}
\caption{Normalized portions of the final spectrum of PKS 0528$-$250,
showing the metal absorptions associated with \zabs = 2.141 DLA
system. The zero velocity corresponds to \zabs = 2.1410, {for the  
\NI\ multiplet we centered on the 1134.4 \AA\ transition}. Smooth lines
are the synthetic spectra obtained from the fit as described in the
text}
\label{f:Q0528-z2.1}
\end{figure*}

%----------------------Fig. 3
\clearpage
 
\begin{figure*}[!h]
\begin{center}
\vspace{0.4cm}
\includegraphics[clip=true,width=11cm,height=12cm]{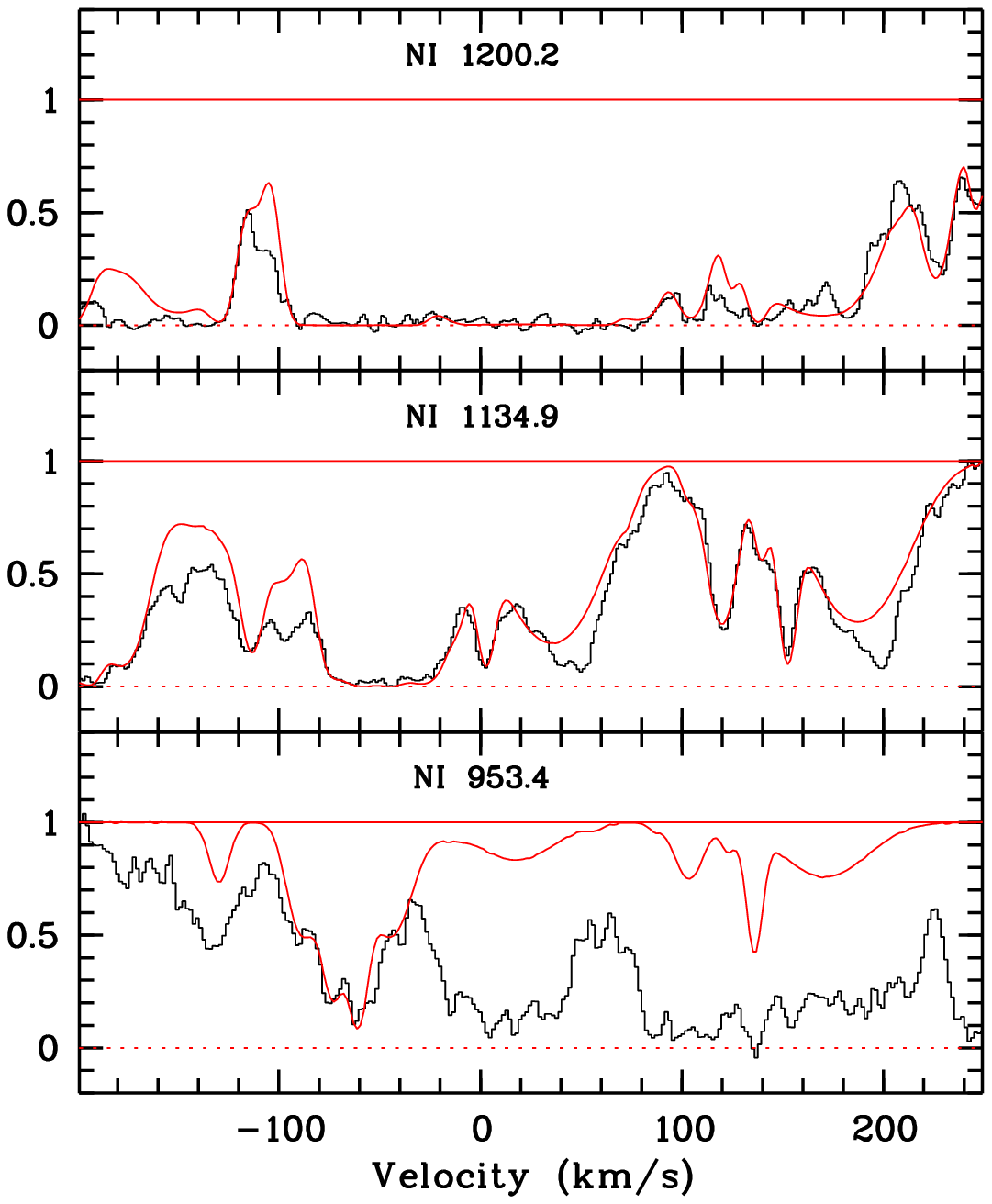}
\end{center}
%\vspace{2cm}
\includegraphics[clip=true,angle=-90,width=8.2cm]{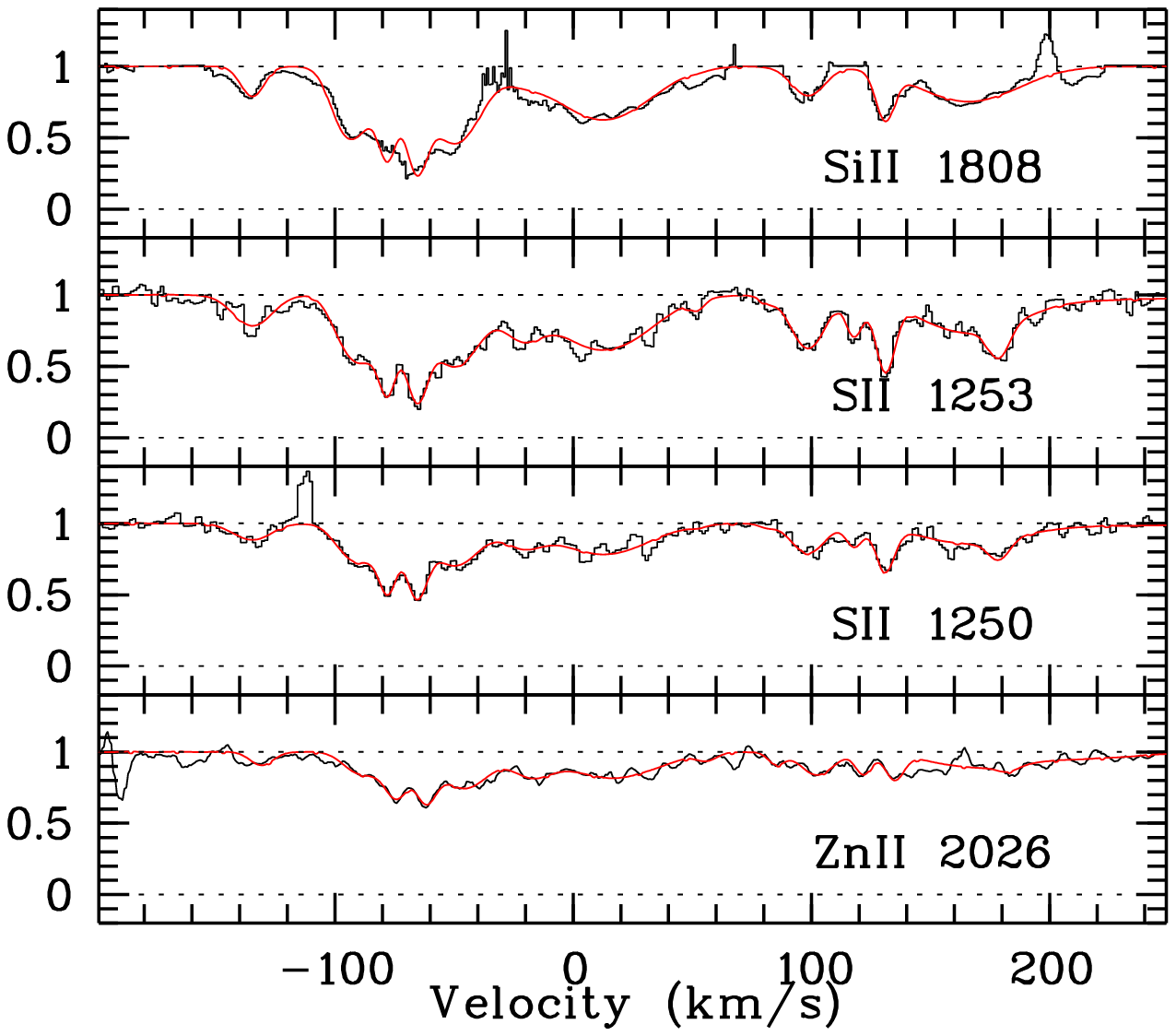}
\hspace{1.2cm}
\includegraphics[clip=true,angle=-90,width=8.2cm]{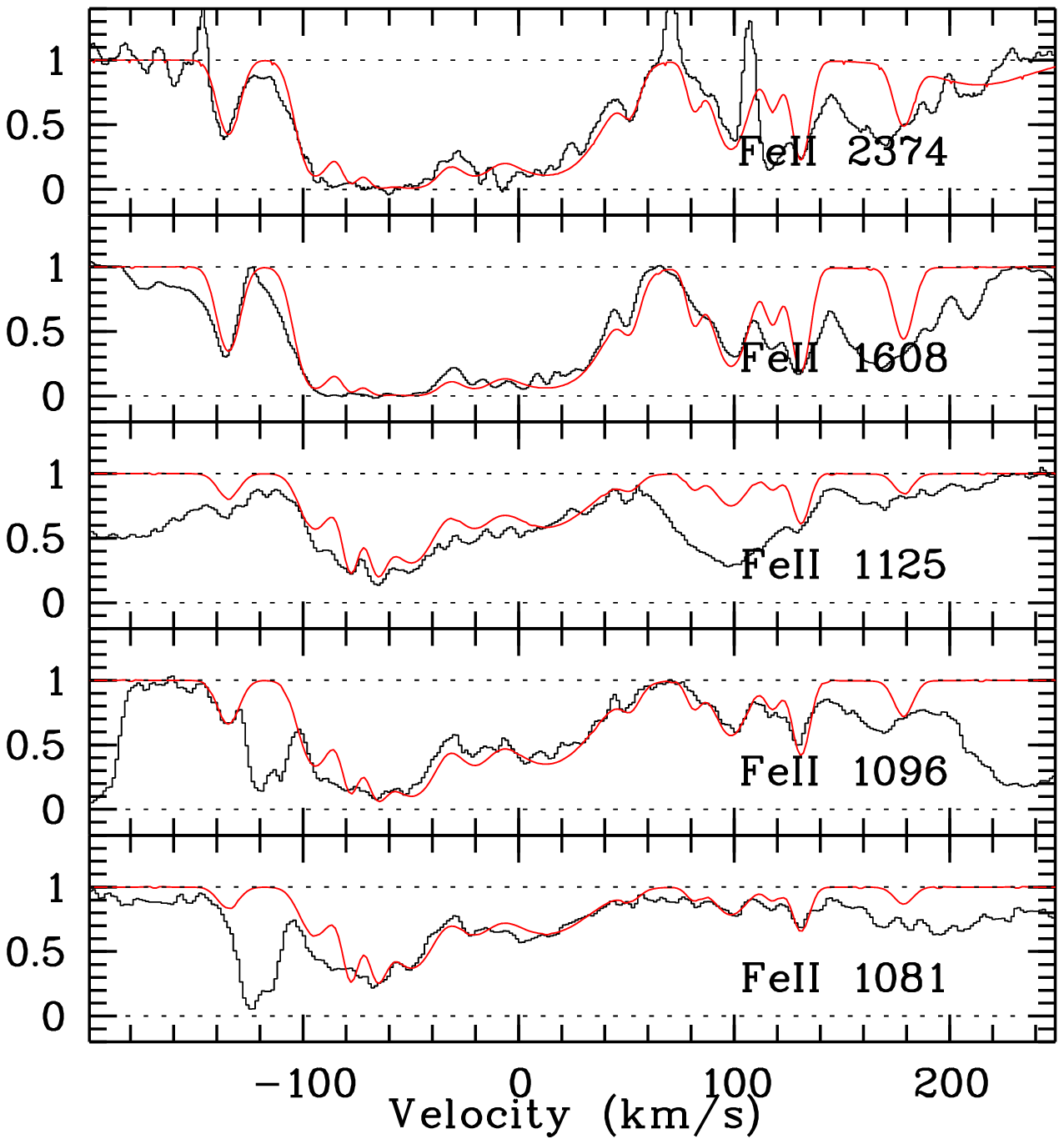}
\caption{Normalized portions of the final spectrum of PKS 0528$-$250,
showing the metal absorptions associated with \zabs = 2.8120 DLA system
(zero velocity).  Smooth lines are the synthetic spectra obtained from
the fit as described in the text}  
\label{f:Q0528-z2.8}
\end{figure*}

%---------------------------Fig. 4
\clearpage
 
\begin{figure*}[!h]
\begin{center}
\vspace{0.4cm}
\includegraphics[clip=true,width=11cm]{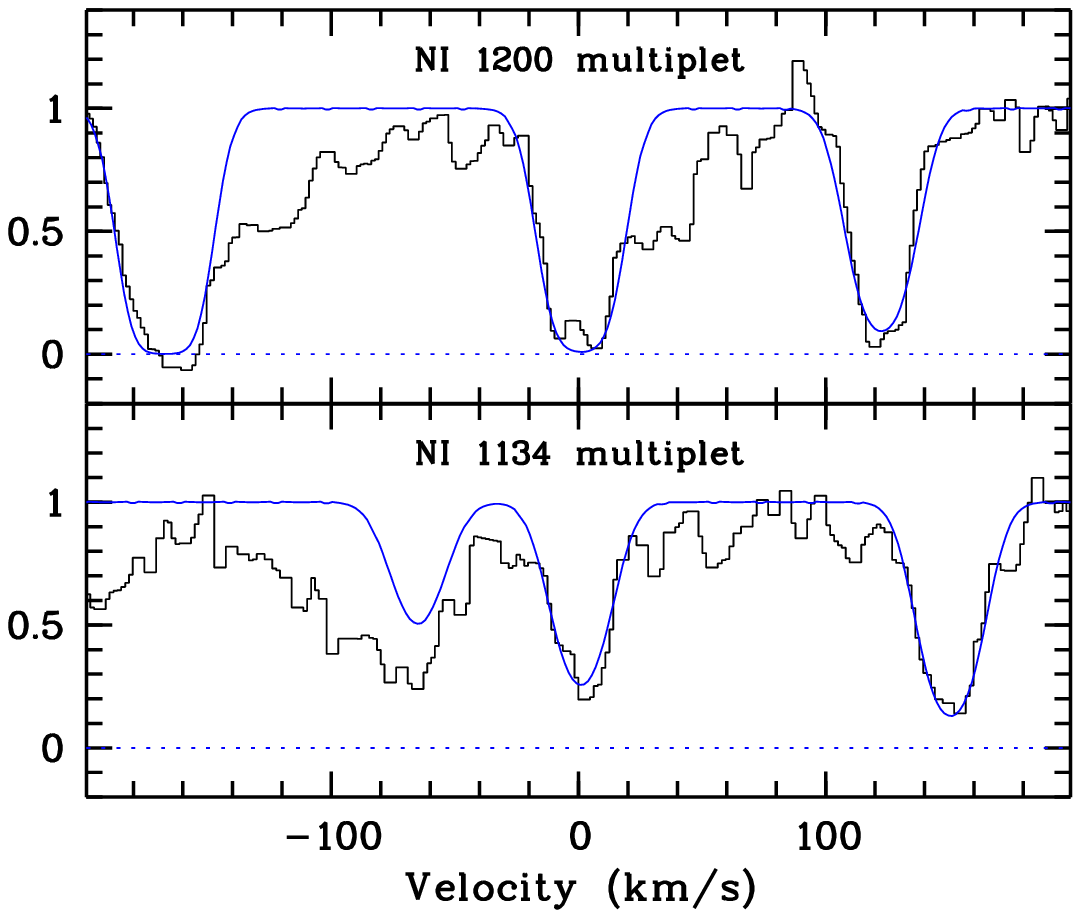}
\end{center}
%\vspace{1cm}
\includegraphics[clip=true,width=8cm,height=15cm]{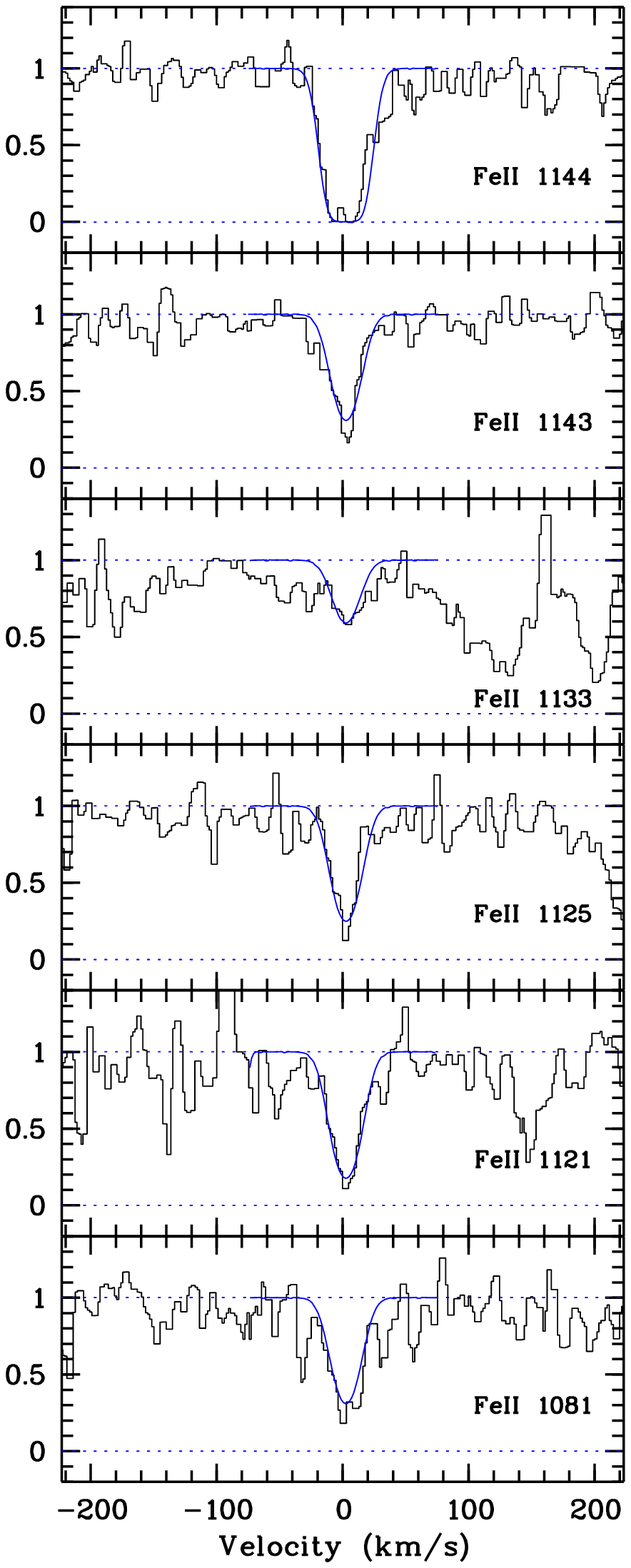}
\hspace{1cm}
\includegraphics[clip=true,width=8cm]{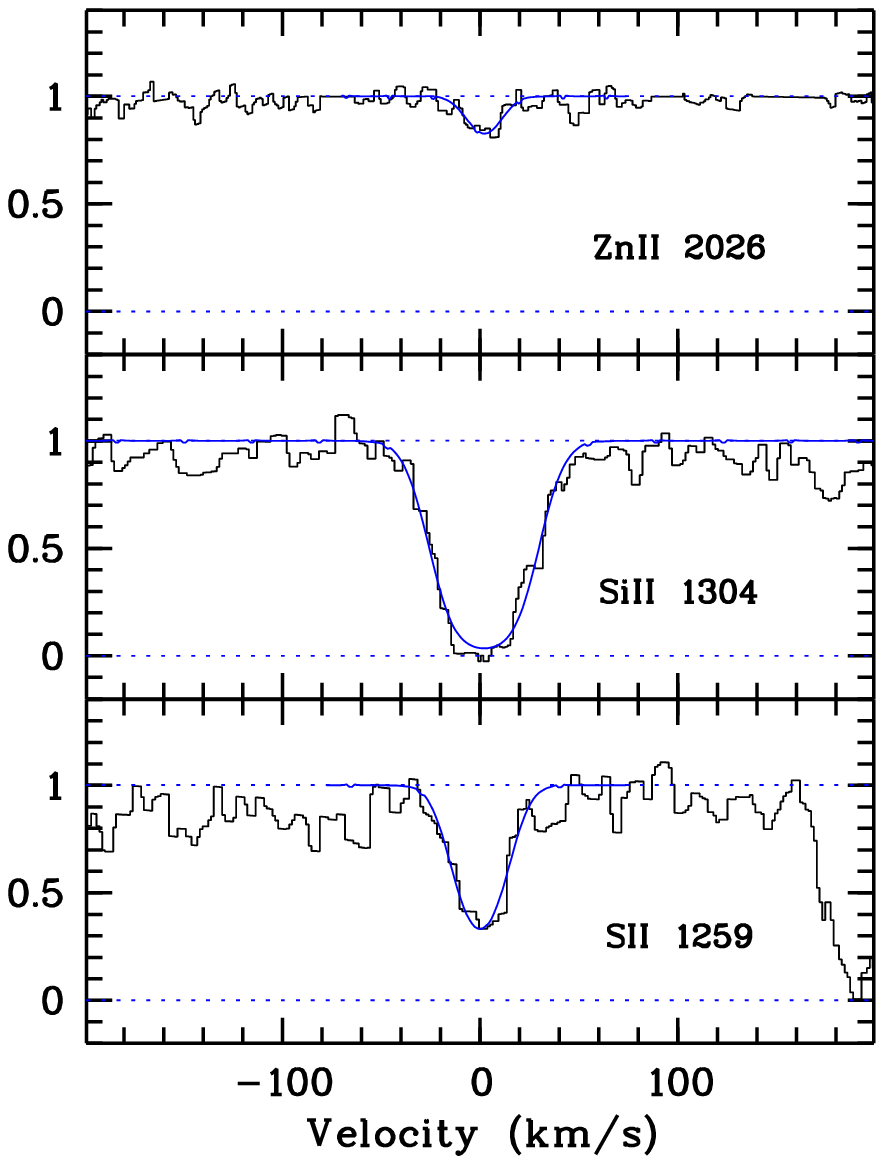}
\caption{Normalized portions of the final spectrum of 
Q0841+126, showing the metal
absorptions associated with \zabs = 2.3745 DLA system. 
The zero velocity has been taken in
correspondence with \zabs = 2.37450, {for  
\NI\ multiplets we centered on the 1134.4 and the 1200.2 \AA\ transitions.} 
Smooth lines are the synthetic spectra obtained from
the fit as described in the text.}
%\label{spectra}
\end{figure*}

%------------------------Fig. 5
\clearpage
 
 \begin{figure*}[!h]
 \begin{center}
 \vspace{0.4cm}
 \includegraphics[clip=true,width=11cm]{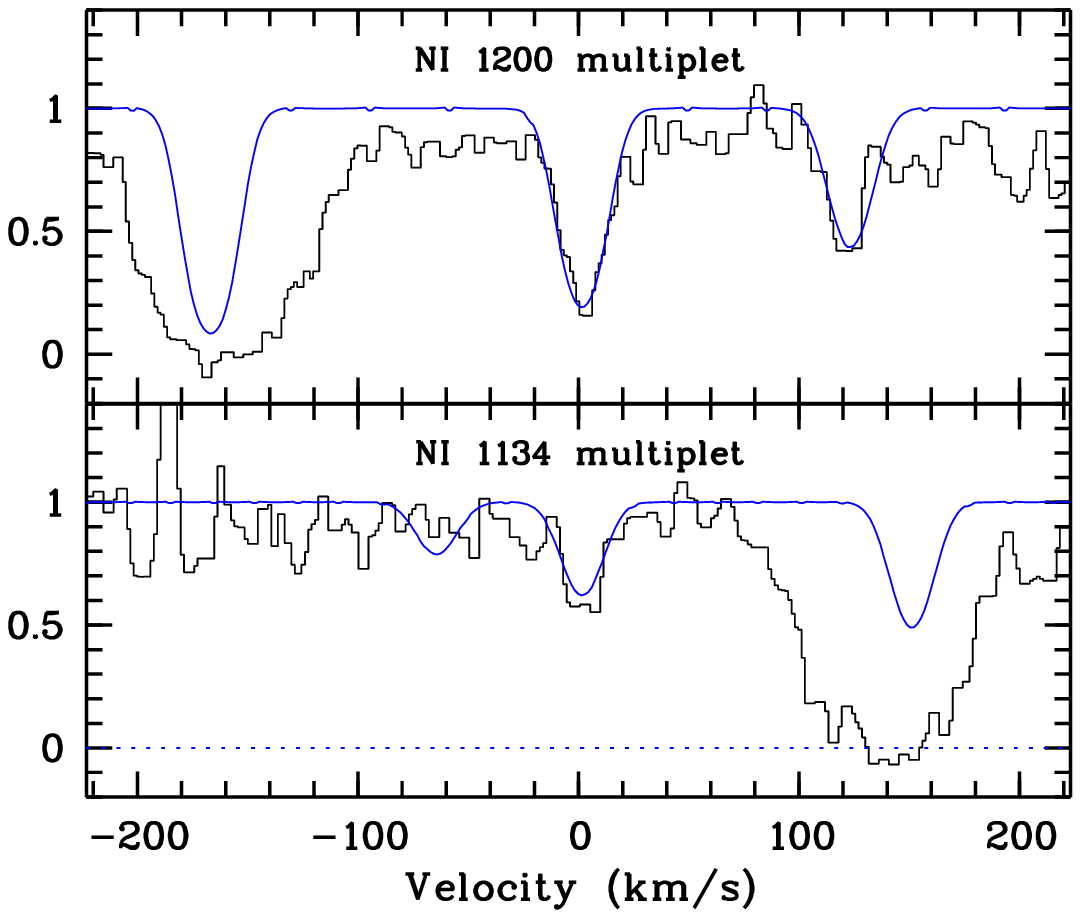}
 \end{center}
 %\vspace{1cm}
 \begin{center}
\includegraphics[clip=true,width=8cm,height=15.5cm]{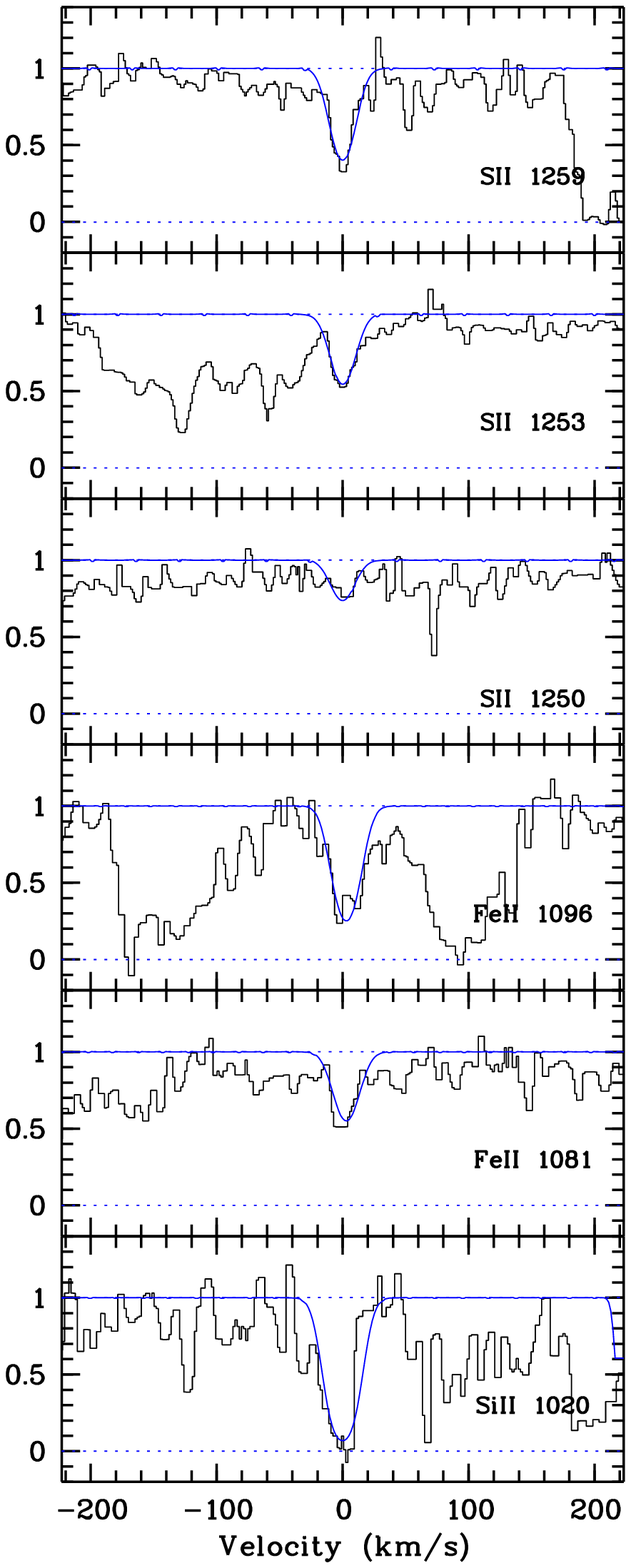}
 \end{center}
 \caption{Normalized portions of the final spectrum of Q0841+126, 
showing the metal
 absorptions associated with \zabs = 2.4762 DLA system (zero velocity), 
{for  
\NI\ multiplets we centered on the 1134.4 and the 1200.2 \AA\ transitions}. 
 Smooth lines are the synthetic spectra obtained from
 the fit as described in the text}
 %%\label{spectra}
  \end{figure*}

%---------------------------------Fig. 6
\clearpage

\begin{figure*}[!h]
\begin{center}
\label{f:Q0940}
\includegraphics[clip=true,angle=-90]{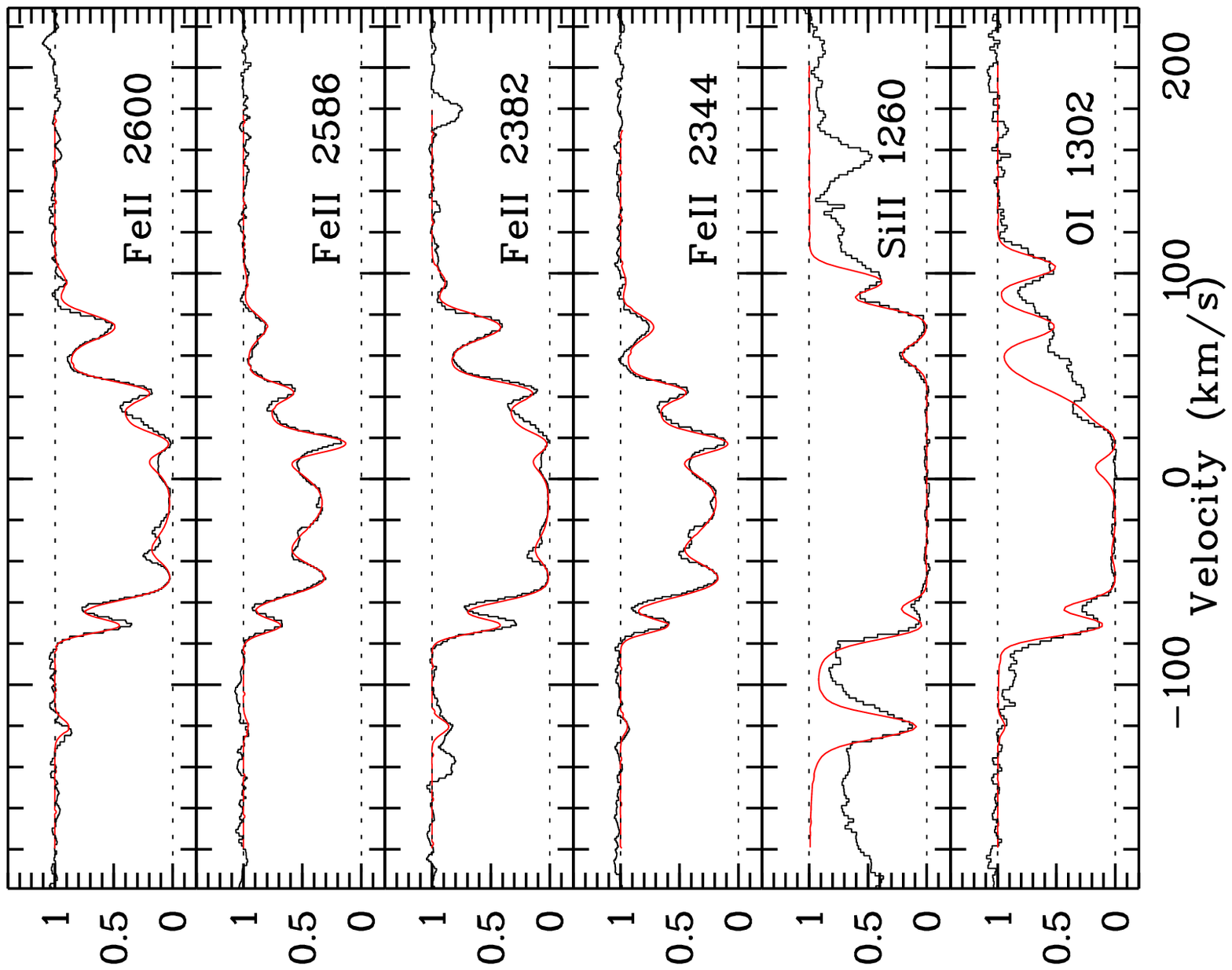}
\end{center}
\caption{Normalized portions of the final spectrum of HE 0940$-$1050,
showing the metal absorptions associated with \zabs = 1.9184 DLA system
(zero velocity).  Smooth lines are the synthetic spectra obtained from
the fit as described in the text}  
%\label{spectra}
\end{figure*}

%---------------------------------Fig. 7
\clearpage

\begin{figure*}[!h]
\begin{center}
\includegraphics[clip=true,width=11cm]{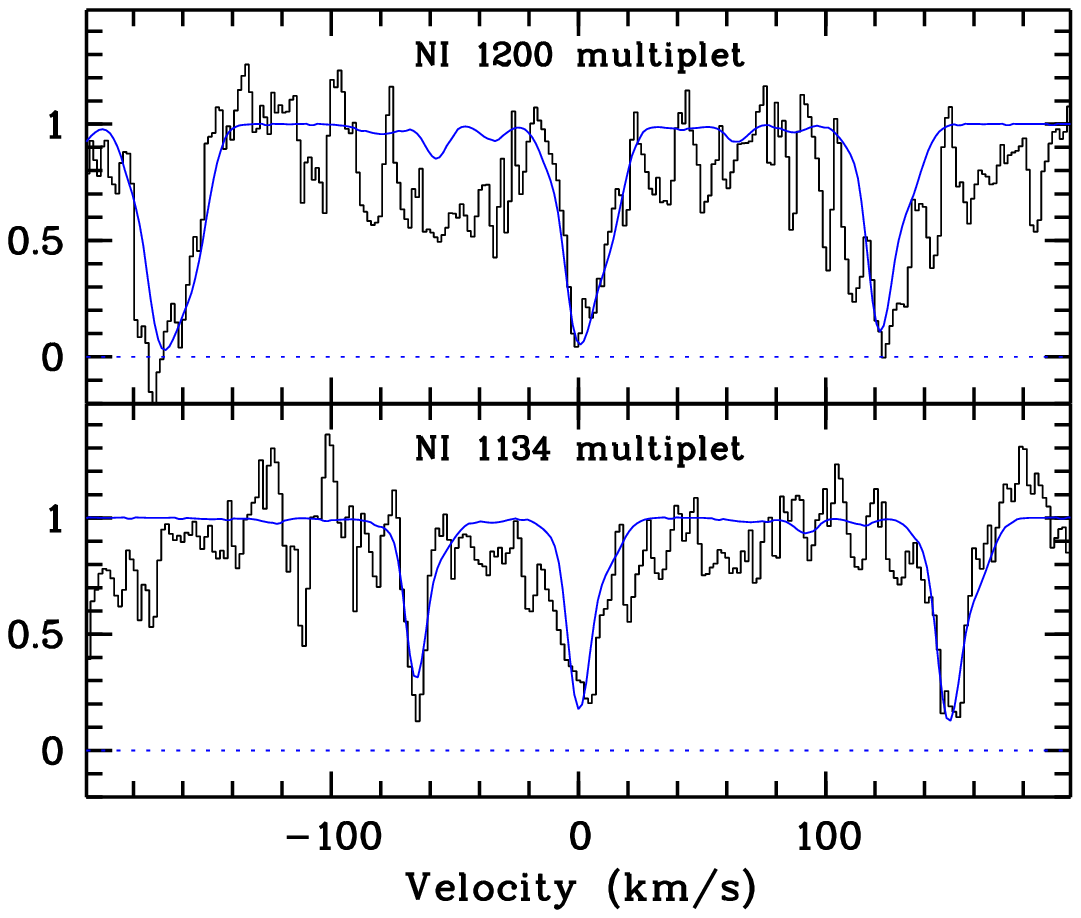}
\end{center}
%\vspace{1cm}
\includegraphics[clip=true,width=8cm]{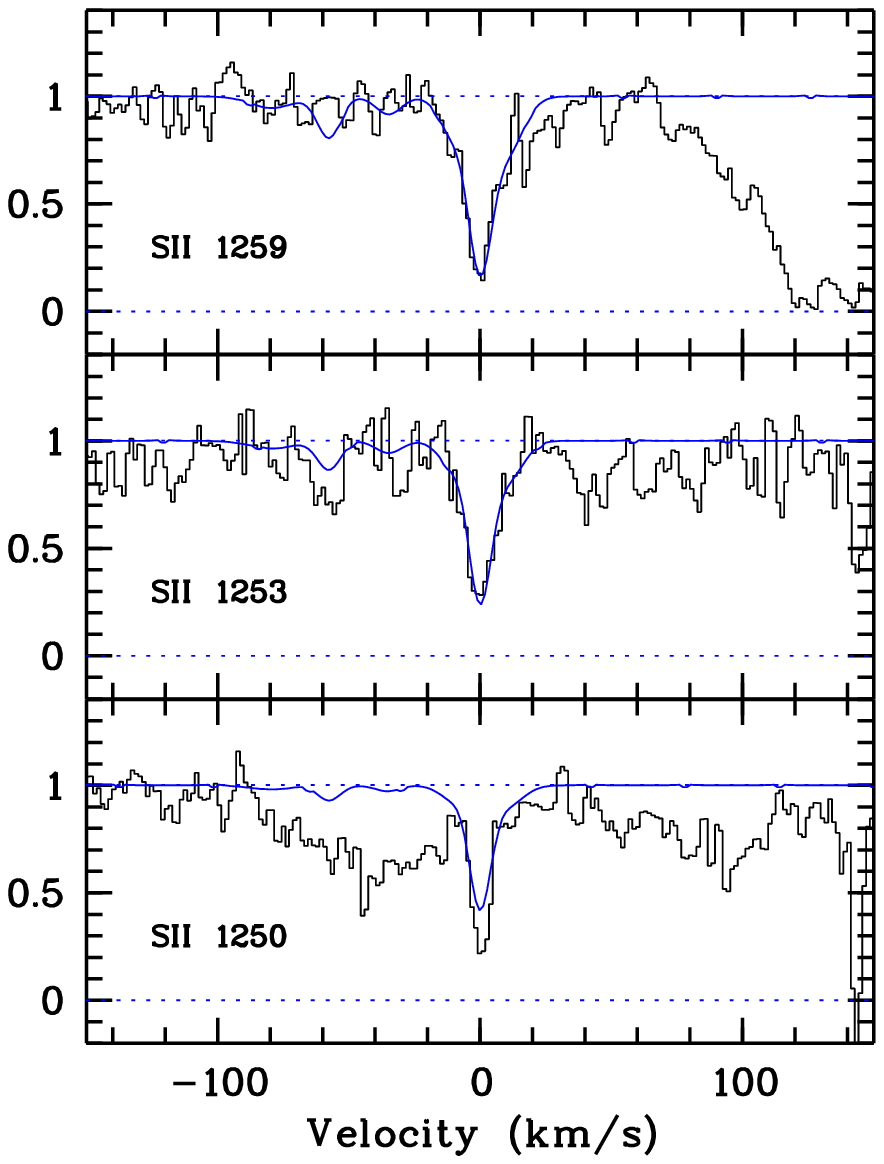}
\hspace{1cm}
\includegraphics[clip=true,width=9cm,height=15cm]{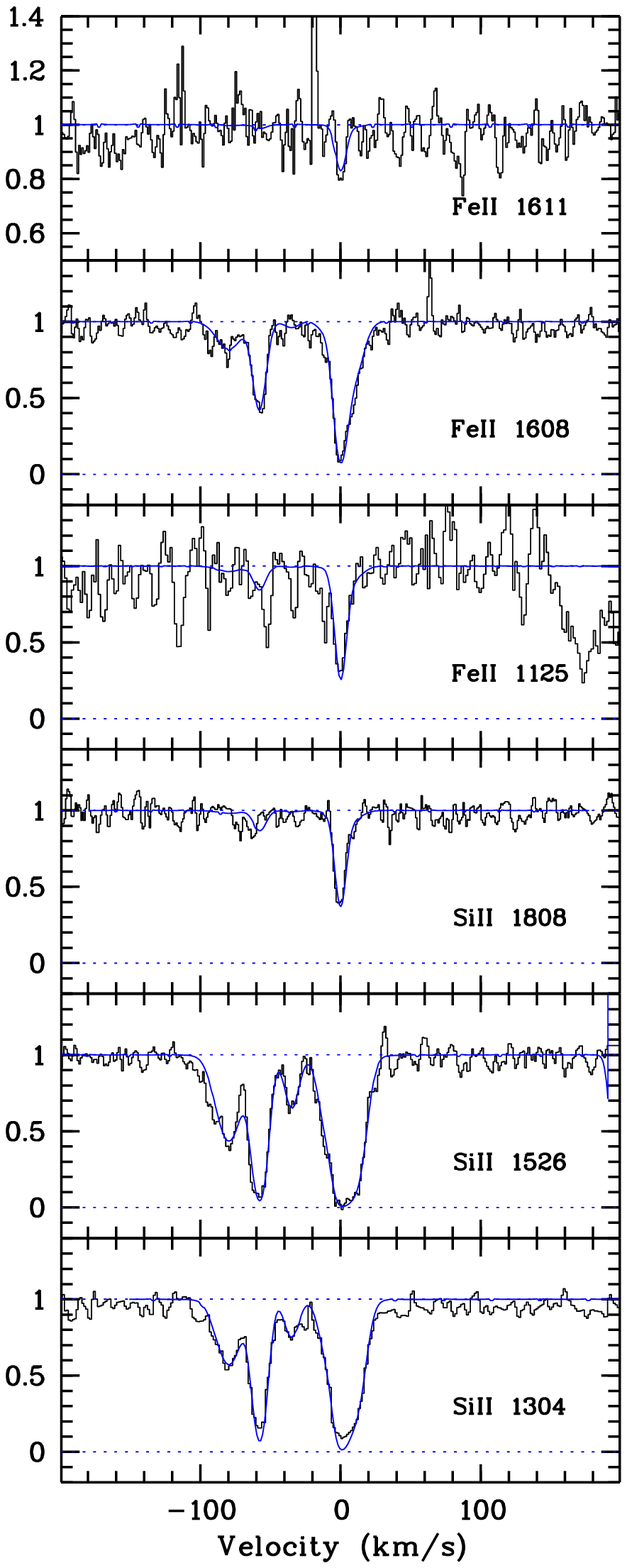}
 \caption{Normalized portions of the final spectrum of 
Q1232+0815, showing the metal
 absorptions associated with \zabs = 2.337 DLA system. The zero velocity 
 corresponds to \zabs = 2.337722, {for  
\NI\ multiplets we centered on the 1134.4 and the 1200.2 \AA\ transitions}.
Smooth lines 
are the synthetic spectra obtained from
 the fit as described in the text}
 %\label{spectra}
 \end{figure*}

%----------------------------Fig. 8
 \clearpage

 \begin{figure*}[!h]
 \begin{center}
 \includegraphics[angle=-90,width=15cm]{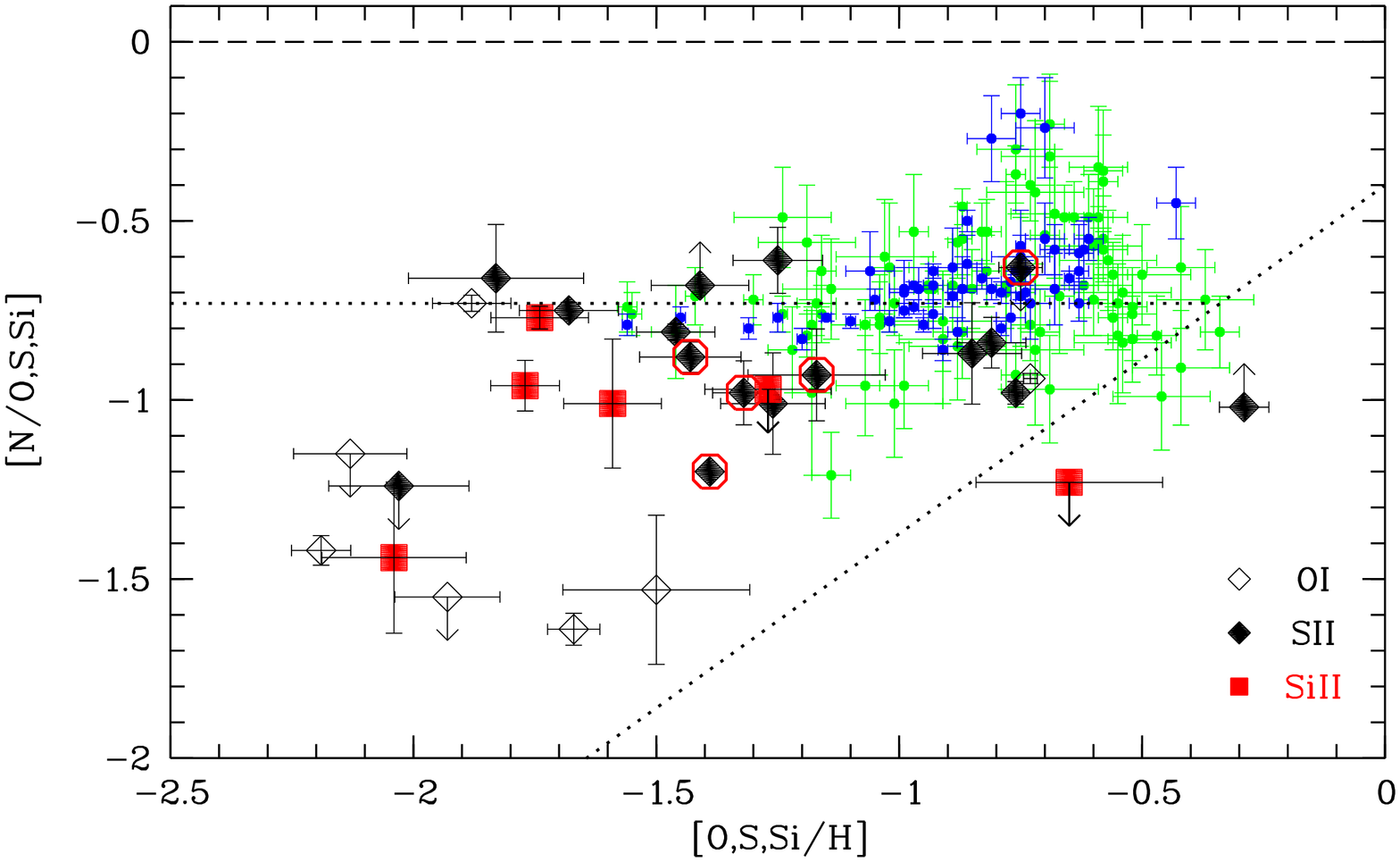}
 \end{center}
 \vspace{0.8cm}
 \begin{center}
 \includegraphics[angle=-90,width=15cm]{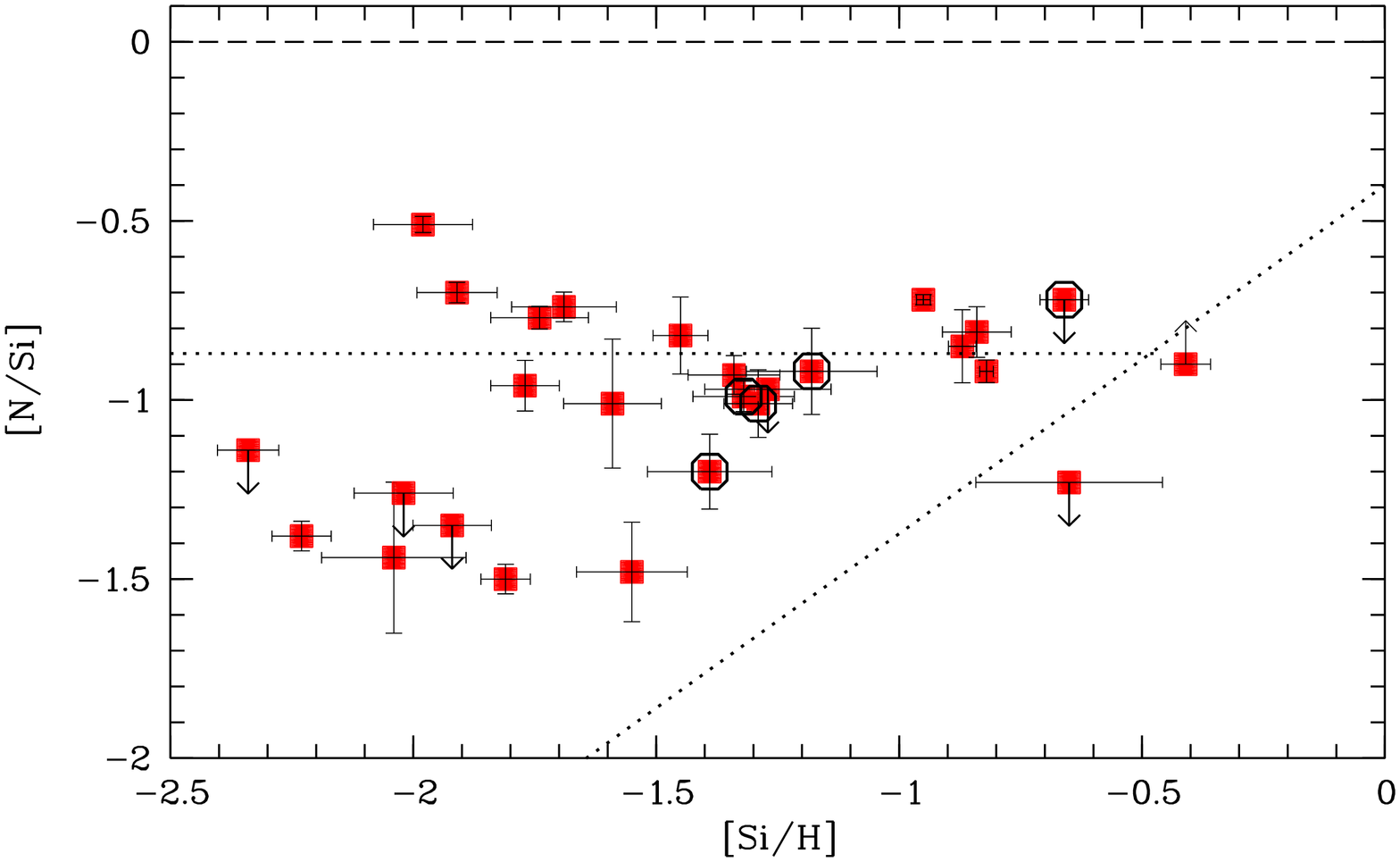}
 \end{center}
 \caption{Upper panel: [N/$\alpha$]  versus metallicity  in DLAs {(different
 big symbols correspond to different $\alpha$-elements, see figure and Table 9 for
 references)}
 and in metal-poor \HII\ regions of dwarf galaxies {(small dots, 
 see text for references). Bold octagons indicate the new DLA measurements 
 presented here}. 
The measurements in 
 BCD galaxies  by Izotov \& Thuan (1999)
are indicated by black dots, while the remaining data by grey dots.
 Dotted lines are empirical
 representations of the {secondary and primary N production. In this panel, the  
 horizontal line (primary) is plotted at the mean value of [N/O] in BCD galaxies.
The solar level is also indicated (dashed line). 
 Lower panel: Same as above but using only the $\alpha$-element silicon in DLAs,
 and their mean [N/Si] to represent the primary N production (horizontal dotted
 line)}}
%%\label{spectra}
  \end{figure*}

%Data are from
%show the [N/O] measurements in \HII\ regions  of dwarfs galaxies   
%(Kobulnicky \& Skillman 1996; van Zee et al.
%1996; van Zee et al. 1997; Izotov \& Thuan 1999).
%The measurements in BCDs by Izotov \& Thuan (1999)
%are indicated by black dots, while the remaining data by grey dots.  

%------------------------------Fig. 9
\clearpage

\begin{figure*}[!h]
 \begin{center}
 \includegraphics[angle=-90,width=15cm]{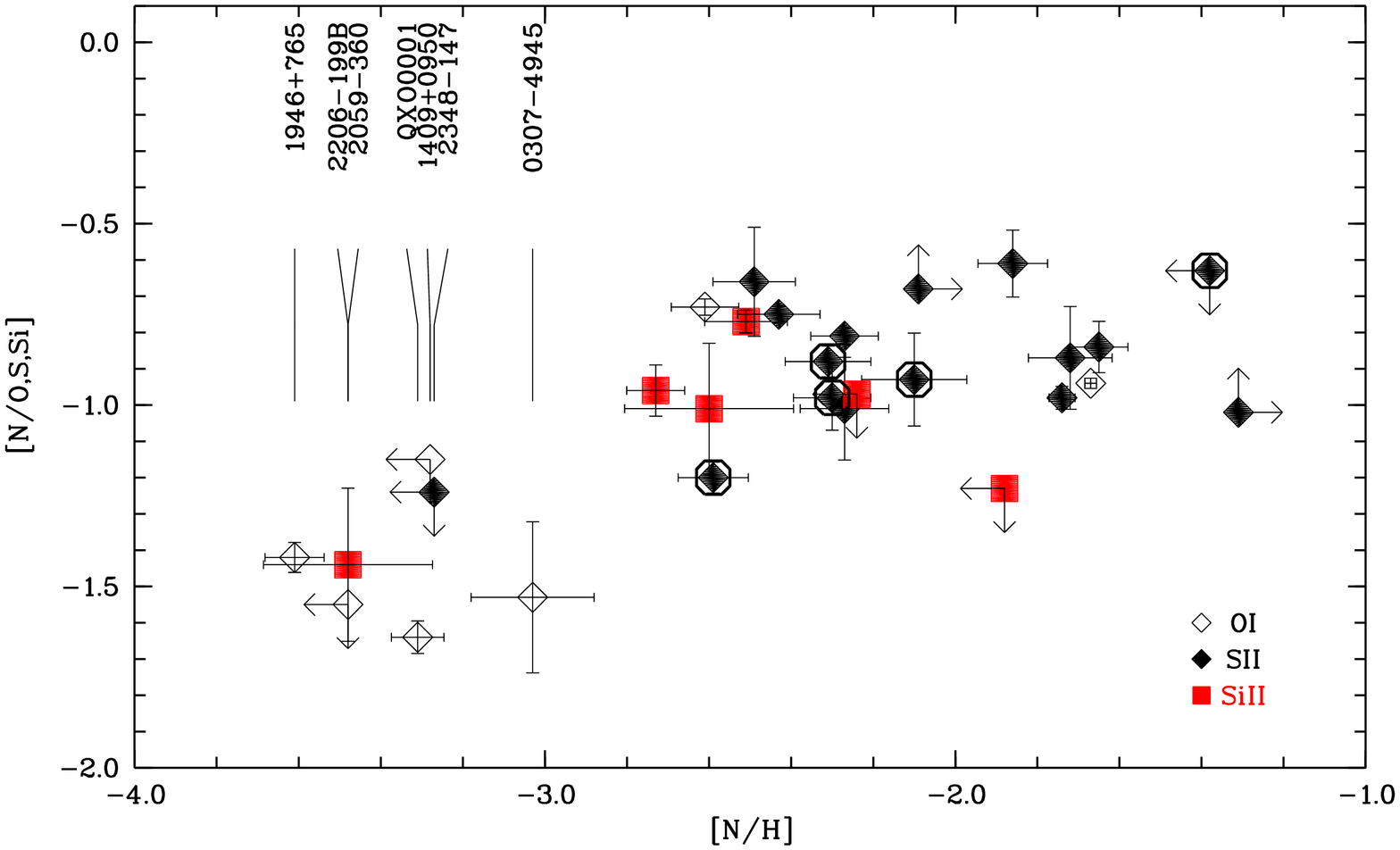}
 \end{center}
 \vspace{0.8cm}
 \begin{center}
 \includegraphics[angle=-90,width=15cm]{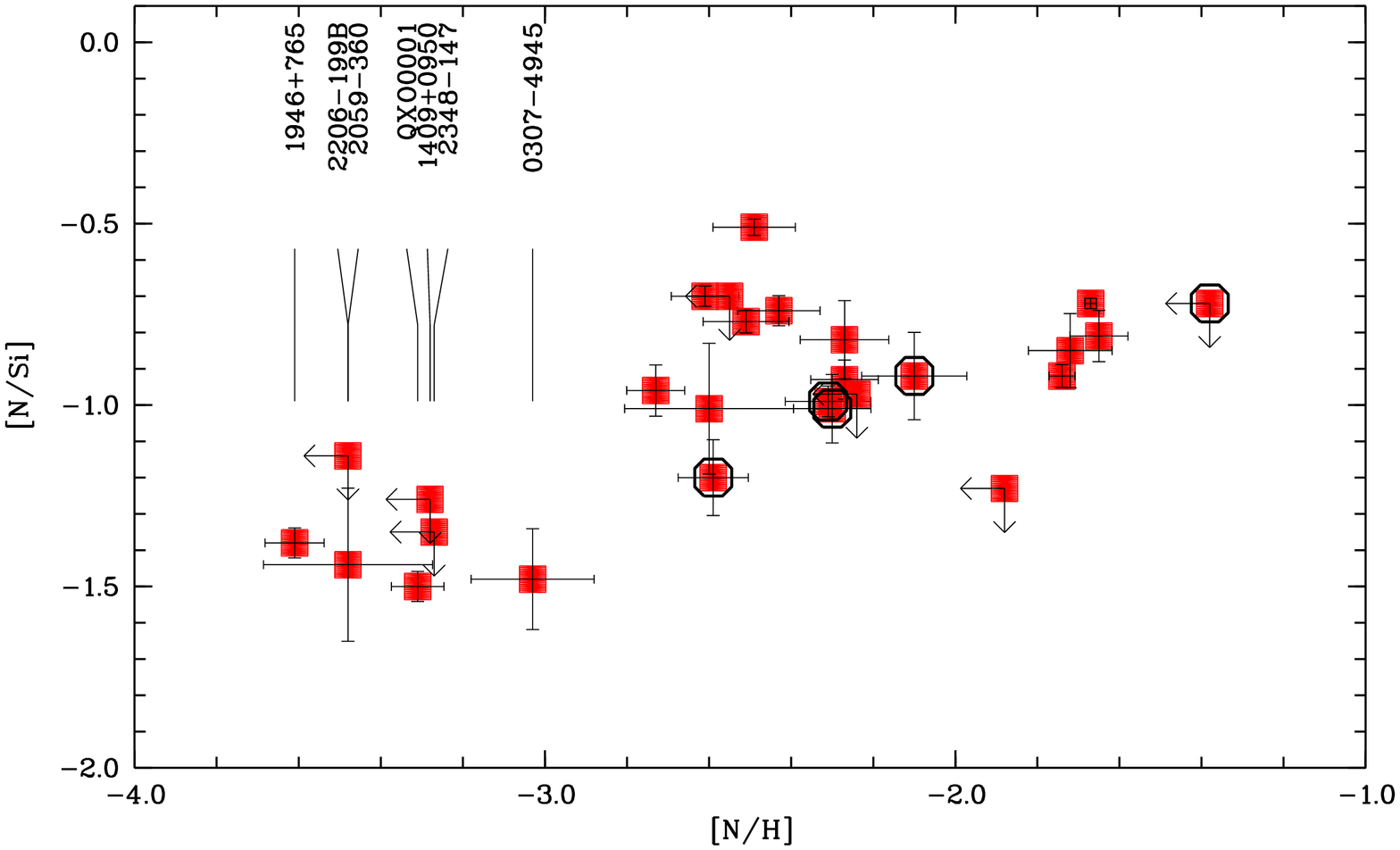}
 \end{center}
\caption{Upper panel: [N/$\alpha$] vs N abundance  in DLAs 
 {(different
 symbols correspond to different $\alpha$-elements, see figure and Table 9 for
 references).  
Bold octagons indicate the new DLA measurements presented here}.
 Lower panel: Same as above but using only the $\alpha$-element silicon }
 %%\label{spectra}
  \end{figure*}

%------------------------------------Fig. 10
\clearpage

\begin{figure*}[!h]
 \begin{center}
 \includegraphics[angle=-90,width=15cm]{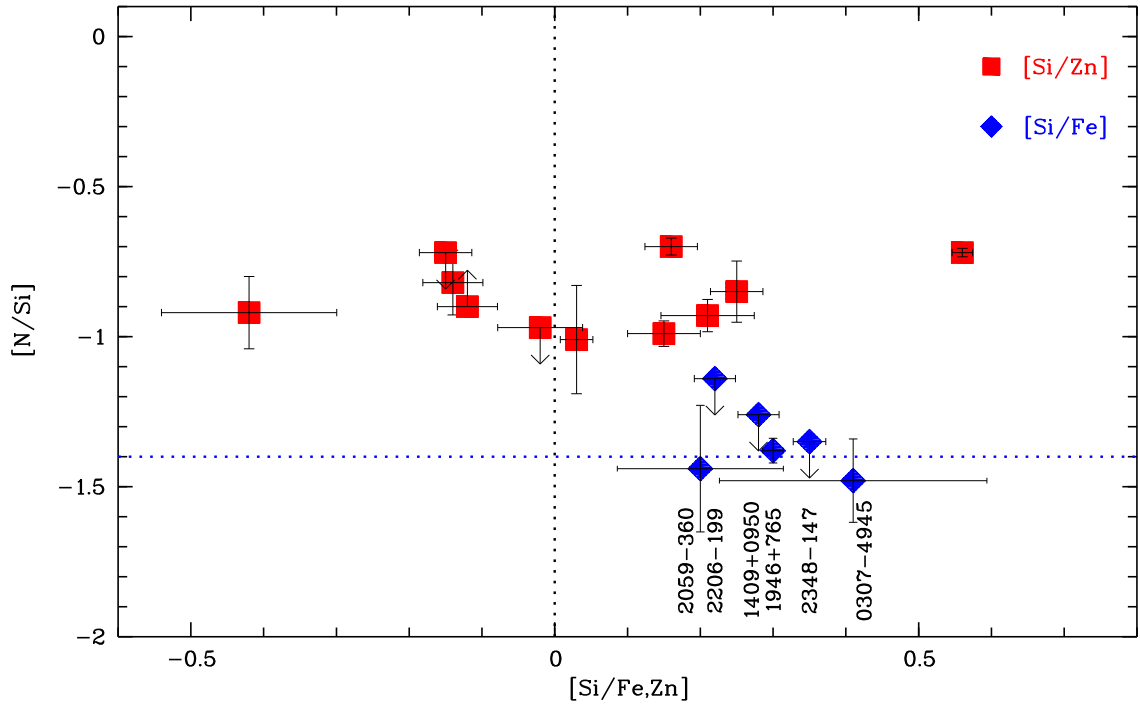}
 \end{center}
 \caption{[N/Si] is plotted versus [Si/Zn] which is only 
available for DLAs with high-N values (squares).
 For DLAs with low-N, labelled in the figure, 
we  plot in abcissa [Si/Fe] ratios since none of these systems have
 determinations of Zn abundance. 
The low-N system towards QXO0001 has been excluded from the plot since it has
 a very poor restrictive lower limit [Si/Fe] $>$ --0.7} 
 \end{figure*} 

%\clearpage

% \begin{figure*}
% \psfig{fig=FeIIAq0841A.eps,width=5cm,height=11cm}
% \psfig{fig=Elementsq0841A.eps,width=5cm,height=11cm}
% \caption{Li doublet for the program stars}
% \label{spectra}
% \end{figure*}

\end{document}